\documentclass[onecolumn,prx, superscriptaddress,longbibliography,margin=1mm 11pt]{revtex4-2} 

\usepackage{ragged2e}
\usepackage{graphicx}
\usepackage{bm}
\usepackage{amsmath}
\usepackage{environ}
\usepackage{appendix}
\usepackage{physics}
\usepackage{rotating,booktabs,array,amsmath,amssymb}
\newcolumntype{C}{>{$\displaystyle}c<{$}} 
\usepackage{caption}
\usepackage{subcaption}

\usepackage[nopar]{lipsum}

\usepackage{enumitem}
\setlist[enumerate,1]{label=(\arabic*),ref=\arabic*}

\makeatletter
\usepackage{amsthm}
\usepackage{amssymb}
\usepackage{amsfonts}
\usepackage{fancyhdr}
\usepackage{slashed}
\usepackage{bbm}
\usepackage{latexsym,epsfig,bbm}
\usepackage[colorlinks=true,urlcolor=blue,citecolor=blue]{hyperref}
\usepackage{todonotes}
\definecolor{blue}{rgb}{0,0.2,1}

\definecolor{red}{rgb}{0.9,0,0}

\newcommand{\vect}[1]{\boldsymbol{#1}}

\begin{document}

\title{Analog quantum simulation of  partial differential equations}
\author{Shi Jin}
\affiliation{Institute of Natural Sciences, Shanghai Jiao Tong University, Shanghai 200240, China}
\affiliation{Ministry of Education Key Laboratory in Scientific and Engineering Computing, Shanghai Jiao Tong University, Shanghai 200240, China}
\affiliation{School of Mathematical Sciences, Shanghai Jiao Tong University, Shanghai, 200240, China}
\affiliation{Shanghai Artificial Intelligence Laboratory, Shanghai, China}
\author{Nana Liu}
\email{nana.liu@quantumlah.org}
\affiliation{Institute of Natural Sciences, Shanghai Jiao Tong University, Shanghai 200240, China}
\affiliation{Ministry of Education Key Laboratory in Scientific and Engineering Computing, Shanghai Jiao Tong University, Shanghai 200240, China}
\affiliation{Shanghai Artificial Intelligence Laboratory, Shanghai, China}
\affiliation{University of Michigan-Shanghai Jiao Tong University Joint Institute, Shanghai 200240, China.}

\date{\today}

\begin{abstract}
Quantum simulators were originally proposed for simulating one partial differential equation (PDE) in particular -- Schr\"odinger's equation. Can quantum simulators also efficiently simulate other PDEs? While most computational methods for PDEs -- both classical and quantum -- are digital (PDEs must be discretised first),  PDEs have continuous degrees of freedom. This suggests that an analog representation can be more natural. While digital quantum degrees of freedom are usually described by qubits, the analog or continuous quantum degrees of freedom can be captured by qumodes. Based on a method called Schr\"odingerisation,  we show how to directly map $D$-dimensional linear PDEs onto a $(D+1)$-qumode quantum system where analog or continuous-variable Hamiltonian simulation on $D+1$ qumodes can be used. This very simple methodology does not require one to discretise PDEs first, and it is not only applicable to linear PDEs but also to some nonlinear PDEs and systems of nonlinear ODEs. We show some examples using this method, including the Liouville equation, heat equation, Fokker-Planck equation, Black-Scholes equations, wave equation and Maxwell's equations. We also devise new protocols for linear PDEs with random coefficients, important in uncertainty quantification, where it is clear how the analog or continuous-variable framework is most natural. This also raises the possibility that some PDEs may be simulated directly on analog quantum systems by using Hamiltonians natural for those quantum systems. 
\end{abstract}

\maketitle


\section{Introduction}
One of the oldest and currently most promising application areas for quantum devices is quantum simulation. Popularised by Feynman in the early 1980s, it is important for more efficient simulation -- compared to its classical counterpart --  of one special partial differential equation (PDE): Schr\"odinger's equation. If quantum simulators can be helpful for simulating Schr\"odinger's equation, it is hoped that they may also be helpful for simulating other PDEs. As with large-scale quantum systems, classical methods for other high-dimensional and large-scale PDEs often suffer from the curse-of-dimensionality, which a quantum treatment might in certain cases be able to mitigate. To enable simulation of PDEs on quantum devices that obey Schr\"odinger's equations, it is crucially important to develop good methods for mapping other PDEs onto Schr\"odinger's equations.\\

Before tackling the question of how to realise this mapping, we first need to consider how  PDEs are represented. Representation of information can be broadly classed into digital, analog or their mixture. Digital information uses discrete and finite degrees of freedom while analog uses continuous and/or infinite degrees of freedom. The simulation of PDEs using any digital representation first requires a discretisation of the PDE. Methods include finite difference, finite element, finite volume, spectral methods, and particle and Monte-Carlo methods. The implementation details and costs are dependent on the discretisation chosen. The majority of classical and quantum methods for PDEs today are based on this paradigm. The digital representation of information in the quantum regime is typically via qubits. Many quantum algorithms for simulating linear PDEs \cite{clader2013preconditioned, childs2021high, costa2019quantum, Javier2022optionprice, linden2020quantum} and systems of nonlinear ODEs and nonlinear PDEs \cite{dodin2021applications,joseph2020koopman,jin2022quantum,jin2023time, liu2021efficient, krovi2023improved, lloyd2020quantum,leyton2008quantum} based on qubits have been theoretically proposed. For PDEs the algorithms can have the complexity $O(\text{poly}(D, \log(1/N)))$ in the best case scenario -- as a quantum subroutine -- where $D$ is the dimension and $N$ is the size of the discretisation needed. \\

\begin{figure}[h] 
\centering
\includegraphics[width=12cm]{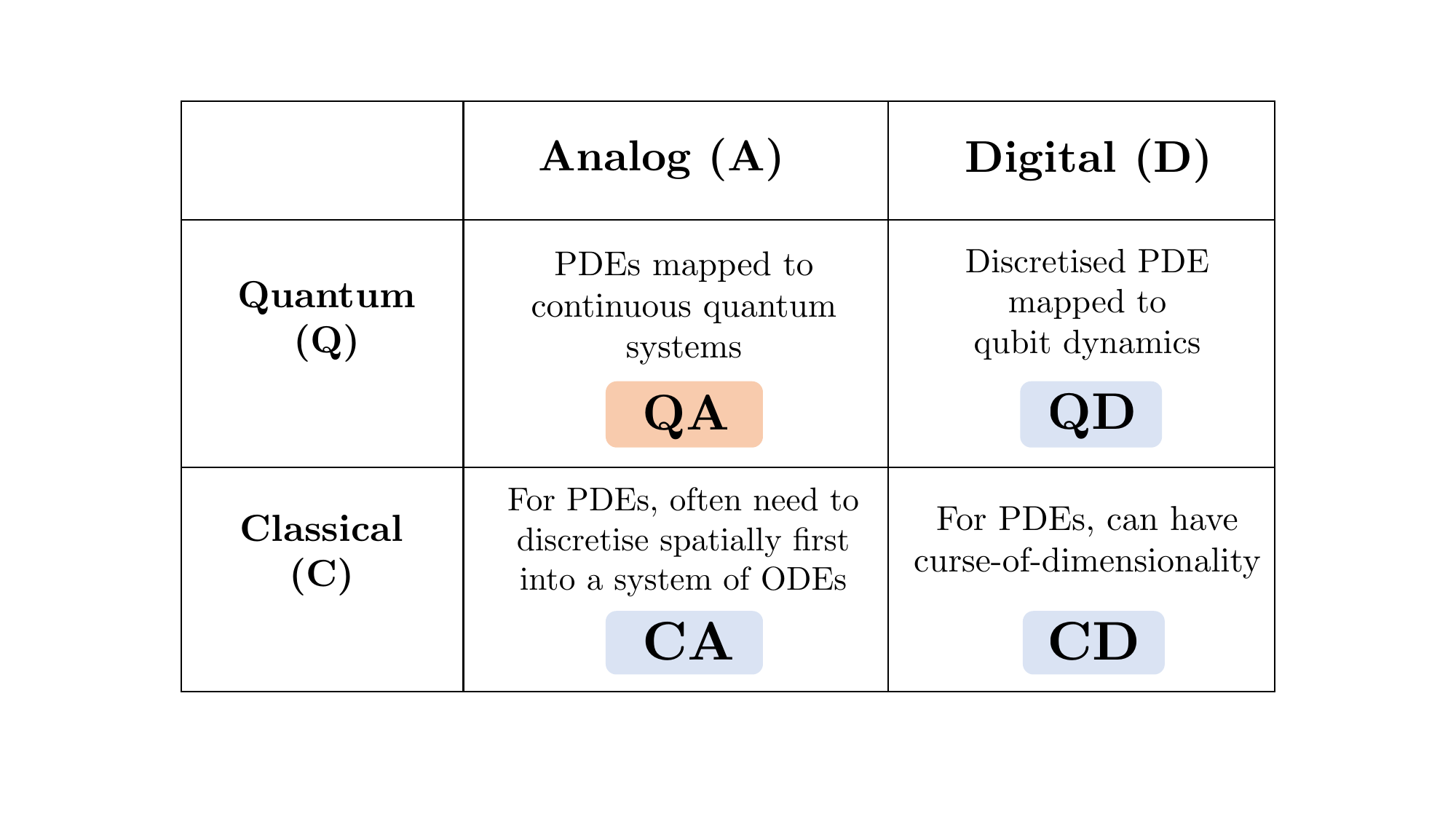} 
\caption{\justifying Table summarising basic possible approaches to the simulation of PDEs. The column `analog' (A) refers to schemes representing the solution with continuous-variables. The column `digital' (D) refers to schemes representing the solution with finite and discrete degrees of freedom. The (QA) regime (orange box) refers to algorithms that simulate the PDEs with the use of continuous-variable quantum systems. There is no need to discretise the PDEs before mapping to the quantum system. This is the main focus of this paper, by using a technique called Schr\"odingerisation, where the evolution in time is also continuous and the corresponding Hamiltonian has a very simple-to-derive form. The unitary generated by the Hamiltonian can either be realised directly by mapping onto an existing quantum system (analogue quantum computing - different notion to analog), or it can be decomposed into a sequence of gates (universal continuous-variable quantum computation). The majority of proposed quantum algorithms for PDEs reside in QD, where discretisation of the PDE is necessary, before it can be mapped onto qubit dynamics. There are also methods in CA, which are primarily for ODEs. To use these for PDEs, one must discretise the PDEs spatially first into a system of ODEs, hence solving PDEs require simultaneously solving many ODEs. The regime CD comprises the majority of algorithms in classical scientific computing, like finite difference methods, etc. Hybrid analog-digital scenarios in this framework are also possible, for example see Sections~\ref{sec:inhomo1}, ~\ref{sec:highert} and~\ref{sec:maxwell}. There are also hybrid classical-quantum schemes, which we will not discuss in this paper. }
\label{tab:4grid}. 
\end{figure}

However, PDEs are intrinsically analog because they typically describe dynamics in space and time which have continuous degrees of freedom. A simple example of solving PDEs in the analog manner, by direct analogy and in common use today, is wind tunnels, for solving Navier-Stokes equations. It is purpose-built for simulating airflow over different geometries and is sufficiently accurate for its purpose. A downside is the difficulty in adapting this, and other specialised devices, to solve more general PDEs and a more universal (though not necessarily fully universal) approach would be preferred. Classical methods for solving ODEs in the analog manner, by indirect analogy, include the use of electronic or mechanical components described by classical physics. Solving PDEs can be difficult with these devices, however, unless they are first discretised into a finite system of ODEs, and the analog machines solve the resulting ODEs \cite{karplus1959analog, jackson1960analog}. To retain the analog nature of PDEs, without first discretising, with the possibility of a more universal approach to simulate large classes of PDEs and with improved efficiency, we can look beyond classical analog systems to quantum systems. See Figure~\ref{tab:4grid} for a simplified categorisation.\\

 Analog information in the quantum regime is commonly known as continuous-variable (CV) quantum information \cite{braunstein2005quantum, lloyd1999quantum}. Many quantum systems, like quantum optical systems and superconducting circuits, are naturally continuous-variable. Indeed, quantum mechanics in the form of wave mechanics was introduced by Schr\"odinger using wave functions, which is a continuous description of information in the state, while the qubit representation was introduced much later. Past methods proposed for CV quantum systems to simulate linear PDEs focus on a CV version of a quantum linear systems solver \cite{arrazola2019quantum}. Methods for ODEs include approximating ODEs through a variational quantum circuit \cite{knudsen2020solving} or using Hamiltonian simulation \cite{aliceCV}. \\
 
 We focus on a new method by directly mapping general linear PDEs evolving in time with $D$ spatial dimensions to a Schr\"odinger-like equation in $D+1$ spatial dimensions. This then enables us to use Hamiltonian simulation in {\it continuous} time. This mapping -- which is {\it exact}  -- is called \textit{Schr\"odingerisation}, recently introduced by the authors in \cite{jin2212quantum, jin2022quantumtechnical}. This method, in contrast to methods based on qubits, enables one to easily identify the necessary Hamiltonian to simulate any given linear PDE, see Figure~\ref{fig:flowchart}. This simplicity also enables an easier way of identifying possible quantum simulators that could be used. Quantum simulators provide the
most plausible near-term device for efficiently simulating a wide range of Hamiltonians \cite{georgescu2014quantum, altman}. The size of the necessary quantum systems are small, requiring only $(D+1)$ qumodes, which are the CV counterparts to qubits. Simulating the same PDE using qubits in the near-term, on the other hand, even for low-dimensional PDEs, would be much more difficult as much larger systems are required when $N$ is large, as system sizes required is $\sim O(D\log N)$. Furthermore, algorithms based on qubits require unitaries that need to be decomposed into smaller gates, rather than relying on a single Hamiltonian that may have the possibility of being naturally realised. While for CV the most general PDEs still require such a decomposition into smaller CV gates, there is still the possibility to find cases where such a decomposition is not necessary. It is also clear from our CV formalism which parts of a given PDE quantum exploit entangling gates and when non-Gaussian operations are required. Both of these are necessary to go beyond efficient classical simulability \cite{bartlett2002efficient}. \\

In Section~\ref{sec:background}, we briefly introduce PDEs and CV quantum systems. In Section~\ref{sec:convection}, we begin with the simplest cases of PDEs to illustrate analog quantum simulation for PDEs without the need of Schr\"odingerisation. For general linear PDEs, we use the Schr\"odingerisation method in Section~\ref{sec:schrodingerisationreview} for homogeneous PDEs, Section~\ref{sec:inhomo1} for inhomogeneous  PDEs and Section~\ref{sec:highert} for PDEs with more than first-order time derivatives. We follow with some examples of important PDEs in Section~\ref{sec:example}, including the Liouville equation, heat equation, the Fokker-Planck equation, the Black-Scholes equation, wave equation and Maxwell's equations (which is a system of PDEs).
We then examine nonlinear PDEs and systems of nonlinear ODEs in Section~\ref{sec:nonlinearsection}.
In Section~\ref{sec:uqpde}, we propose a new quantum algorithm for uncertain linear PDEs, based on stochastic Galerkin methods. Uncertain PDEs play an important role in the subject of uncertainty quantification. For these algorithms it is particularly clear that the CV or analog representation is the most natural and elegant.  \\

We summarise our findings in Table~\ref{tab:summary}. 

\section{Background} \label{sec:background}

\subsection{Partial differential equations}

The most general PDEs for $u(t,x_1,...,x_D)$ in $D$ spatial dimensions and first-order in time has the form
\begin{align}\label{eq:generalpde}
    \frac{\partial u}{\partial t}+F(u, \nabla u, \nabla^2u,...,\nabla^n u)=0, \qquad t\geq 0, \qquad x_j \in \mathbb{R}, \qquad j=1,...,D,
    \end{align}
where if $F$ is a linear function, then the above is a linear PDE. If $F$ is a nonlinear function of any of its arguments $u, \nabla u,$...etc, then the above is a nonlinear PDE.  If all the terms in Eq.~\eqref{eq:generalpde} involve $u$, then it is called a homogeneous PDE. If there are also terms of the kind $f(x_1,...,x_D)\neq 0$, independent of $u$, then it is an inhomogeneous PDE.\\

For most PDEs, whether linear or nonlinear, analytical solutions seldom exist. This means that most of the time one resorts to numerical algorithms that are implemented on classical (digital) computers. These algorithms are among the most important achievements in scientific and engineering computing. Different types of partial differential equations have drastically different behavior in their solutions. For example, they could differ in regularity (smoothness) or have different physical properties (e.g. positivity, conservation, invariance, entropy conditions,... etc). Thus different PDEs require drastically different numerical strategies. \\

However, all these methods generally involve some kind of discretisation of spatial degrees of freedom. For general PDEs, one can write it as a system of ODEs by first discretising in space. There are two classes of discretisation methods. One is the so-called Eulerian framework, in which one discretises the spatial variables by a finite difference, finite element, finite volume, or spectral method, on a fixed mesh. The advantage of Eulerian methods is their high order accuracy but they suffer from the curse-of-dimensionality in high spatial dimensions. The other is the Lagrangian method, including the so-called particle, and mesh-freemethods. This is popular for high dimensional problems, for instance particle or Monte-Carlo methods for kinetic equations (the Boltzmann equation and Vlasov type equations), and vortex methods for incompressible Euler and Navier-Stokes equations in fluid dynamics. These Lagrangian methods do not suffer from the curse-of-dimensionality but they are of lower order (typically only half to
first order) methods, and also become inaccurate in areas where there are not enough particles to accurately approximate, for example, field equations (like in a Particle-in-Cell simulation of Vlasov-Poisson or Vlasov-Maxwell equations in plasma).\\

The curse-of-dimensionality presents a major obstacle for numerical methods for PDEs, and it arises chiefly out of the need to perform spatial discretisation in the Eulerian framework. This is problematic especially for systems with large dimensions (like $n$-body Schr\"odinger equations, kinetic equations such as the Boltzmann equation, Landau and radiative transfer equations, and PDEs with uncertainties) or with requirement of high resolutions (for example problems with small scales and high frequencies). \\

To resolve this issue, quantum algorithms have been proposed on an analogous problem, which is the simulation of quantum states whose amplitudes contain the classical solutions. There is a large body of literature on this topic where the bulk of these quantum methods focus on representing the discretised solution in the form of {\it qubits}. For instance, consider a problem in $D$ dimensions to be solved with $N$ mesh points per space dimension, then there are $N^D$ points to be used, which can be embedded into $d\log_2 N$ qubits. Then, when certain conditions hold (see references in the introduction), it can be shown that the quantum cost can be polynomial in $D$ by using methods mostly based on quantum linear algebra, thus greatly alleviating the curse-of-dimensionality problem. \\

However, very few methods to date have taken advantage of existing {\it continuous-variable} degrees of freedom in physical systems to simulate PDEs. By using continuous-variable systems to perform these simulations, no discretisation is actually necessary, at least in the ideal scenario. We will briefly introduce quantum continuous-variable systems in the next section and later embed both linear and nonlinear PDEs in these systems.

\subsection{Continuous-variable quantum systems}
\label{sec:cvintro}

Quantum information processing protocols are often introduced in terms of qubits, which are the quantum counterparts to classical bit-strings $\{0,1\}^{\otimes m}$. A qubit uses the eigenbasis $\{|0\rangle, |1\rangle\}$ for the computational basis. An $m$-qubit system is a tensor product of $m$ qubits which lives in $2^m$-dimensional Hilbert space. \\

A continuous-variable (CV) quantum state, or {\it `qumode'}, on the other hand, spans an infinite-dimensional Hilbert space. A qumode is the quantum analogue of a continuous classical degree of freedom, like position, momentum or energy before being quantised. A qumode is equipped with observables with a continuous spectrum, such as the position $\hat{x}$ and momentum $\hat{p}$ observables of a quantum particle. Its eigenbasis can be chosen to be for instance $\{|x\rangle\}_{x \in \mathbb{R}}$, which are the eigenstates of $\hat{x}$. In this basis, a qumode can be expressed as $|u\rangle=(1/\|\vect{u}\|)\int u(x)|x\rangle dx$, where $\|\vect{u}\|^2=\int dx |u(x)|^2$ is the normalisation constant. A system of $m$-qumodes is a tensor product of $m$ qumodes. See \cite{navarrete2016introduction} for a good  introduction on the CV formalism and how it can be used in quantum information.\\

Despite the relatively less well-known terminology of qumodes, physicists will no doubt observe that $u(x)/\|\vect{u}\|$ is a familiar quantity as it is precisely the wavefunction of the quantum state in the position basis. Thus, despite CV quantum information processing and computation being less well-known compared to processing with qubits, it is clear that these are the natural states of quantum systems. Extra effort is generally required to reduce the wavefunction or qumode description of quantum systems in the laboratory to that of qubits.\\

The quadrature operators of a qumode are $\hat{x}$ and $\hat{p}$, where $[\hat{x},\hat{p}]=i$. If we let $|x\rangle$ and $|p\rangle$ denote the eigenvectors of $\hat{x}$ and $\hat{p}$ respectively, then $\langle x|p\rangle=\exp(ixp)/\sqrt{2\pi}$. The position and momentum eigenstates each form a complete eigenbasis so $\int dx |x\rangle \langle x|=I=\int dp |p\rangle \langle p|$. Here the quantised momentum operator $\hat{p}$ is also associated with the spatial derivative $\hat{p} \leftrightarrow -i\partial/\partial x$. One can see this easily from the same commutation relation being obeyed $[x, -i\partial/\partial x]u=iu$. Suppose we define the state $|\partial u/\partial x\rangle=\int  (\partial u/\partial x)|x\rangle dx$. Then it is straightforward to show $i\hat{p}|u\rangle=|\partial u/\partial x\rangle$. For simplicity, we ignore normalisation constants and it is simple to prove the following. Let $\tilde{u}(p)$ denote the Fourier transform of $u(x)$. Then $i\hat{p}|u\rangle=i\int u(x)\hat{p}|x\rangle dx =i\iint u(x)\hat{p}|p\rangle \langle p|x\rangle dx dp =i \iint  u(x)p\exp(-ixp)|p\rangle dx dp=i\int \tilde{u}(p)p|p\rangle dp=i \iint \tilde{u}(p)p |x'\rangle \langle x'|p\rangle dp dx'=i\iint \tilde{u}(p)p \exp(ix'p)|x'\rangle dx'dp=\iint (\partial/\partial x')(\tilde{u}(p)\exp(ix'p))|x'\rangle dx'dp=\int \partial u(x')/\partial x'|x'\rangle dx'=|\partial u/\partial x\rangle$. Similarly $(i\hat{p})^n|u\rangle=|\partial^n u/\partial x^n\rangle$. This simple observation will be important for our simulation of PDEs later. \\

These quadrature operators can also be written in terms of creation and annihiliation operators $\hat{a}, \hat{a}^{\dagger}$. These are related to the quadrature operators by $\hat{x}=(\hat{a}+\hat{a}^{\dagger})/\sqrt{2}$ and $\hat{p}=(\hat{a}-\hat{a}^{\dagger})/(\sqrt{2}i)$. The number operator is defined as $\hat{n}=\hat{a}^{\dagger}\hat{a}$ with corresponding eigenvector $|n\rangle$ and eigenvalue $n$, where the latter is interpreted as particle number. The eigenbasis $\{|n\rangle\}_{n=0}^{\infty}$ forms a so-called number state or Fock state basis set. In this notation, $|0\rangle$ is the vacuum state.\\


Consider a system of $m$-qumodes, then we define the position/momentum operator only acting on the $j^{\text{th}}$ mode as 
 \begin{align}
        \hat{p}_j=I^{\otimes j-1}\otimes \hat{p} \otimes I^{\otimes m-j}, \qquad \hat{x}_j=I^{\otimes j-1}\otimes \hat{x} \otimes I^{\otimes m-j}, \qquad [\hat{x}_j,\hat{p}_k]=i\delta_{jk}.
    \end{align}

Measurement of the quadrature operators is also routinely performed, for instance through homodyne detection. Mathematically, this corresponds to a projective measurement of a quadrature operator. In fact, the expectation value of any linear combination of quadratures $\begin{pmatrix}
    \hat{x}_{\theta} \\
    \hat{p}_{\theta}
\end{pmatrix}=\begin{pmatrix}
    \cos(\theta) & \sin(\theta) \\
    -\sin(\theta) & \cos(\theta)
\end{pmatrix} \begin{pmatrix}
    \hat{x}_{0}\\
    \hat{p}_{0}
\end{pmatrix}$, $\theta \in [0, \pi/2)$ can be measured using homodyne detection in a rotated basis. \\

Operations on qumodes fall into two classes: Gaussian and non-Gaussian. A universal gate set for qumodes can be formed from a set of Gaussian operations and only a single kind of non-Gaussian operation. This is analogous to how a universal gate set for qubits consist of Clifford and non-Clifford gates, where Clifford gates are the analogy to Gaussian gates and non-Clifford gates are analogous to non-Gaussian operations. A pure qumode Gaussian state can be expressed as $U_G|0\rangle$ where $U_G$ is the Gaussian unitary operation characterised by the affine map $(\vect{S}, \vect{d}):(\hat{x}, \hat{p})^T \rightarrow \vect{S}(\hat{x}, \hat{p})^T+\vect{d}$ for $\vect{S} \in \mathbb{R}^{2 \times k}$, $\vect{d} \in \mathbb{R}^{2}$. For example, a coherent state $|\alpha\rangle$ is a Gaussian state prepared by a Gaussian operation $U_G(\alpha)=\exp(\alpha \hat{a}^{\dagger}-\alpha^*\hat{a})=\exp(i\sqrt{2}(\hat{x}\text{Im}(\alpha)-\hat{p}\text{Re}(\alpha)))$. The Hamiltonian generating this $U_G$ only contains terms that are linear in the quadratures. Gaussian operations also contain operations whose generating Hamiltonians are quadratic in the quadrature operations. For quadratic operations on a single mode, examples include single-mode squeezing operators. For quadratic operations across two-modes, an example is the so-called CZ gate $U_G=\exp(i\hat{x} \otimes \hat{x})$. For operations across two-modes like the CZ gate, we note that there is entanglement generation. For any order higher than two, we refer to these as non-Gaussian operations. \\

While Gaussian operations are generally quite accessible in laboratory settings, non-Gaussian operations are less so. However, they are known to be necessary for universal CV computation \cite{lloyd1999quantum} and also necessary to go beyond efficient classical simulability \cite{bartlett2002efficient}.  While concatenations of Gaussian operations only produce other Gaussian operations, only a single kind of non-Gaussian operation and Gaussian operations are sufficient to produce any other non-Gaussian operation. Thus, Gaussian operations with a single kind of non-Gaussian gate forms a universal gate set for qumodes. A typical non-Gaussian gate has the form $U_{nG}=\exp(it\hat{x}^r)$, $r\geq 3$. Using methods similar to Trotterisation --  employing commutation relations of CV operators $[\hat{x}, \hat{p}]=i$ -- any CV unitary can be decomposed into a finite number of CV gates from the universal set \cite{sefi2011decompose}. \\

In this paper, we use $[\hat{A}, \hat{B}]=\hat{A}\hat{B}-\hat{B}\hat{A}$ to denote commutation relations and $\{\hat{A}, \hat{B}\}=\hat{A}\hat{B}+\hat{B}\hat{A}$ to denote anticommutation relations.

\subsubsection*{Physical platforms} \label{sec:lab}

In this paper, we show how any PDE can be simulated using qumodes through finding the appropriate generating Hamiltonian $\vect{H}$. We will find that for some PDEs, only Gaussian operations are necessary while for others non-Gaussian operations are essential. We also show how entangling gates like CZ gates are necessary for most PDEs. These unitaries can be generated in one of two main ways. The first way is by breaking the unitary up into the set of gates in the CV universal gate set (with a set of Gaussian gates and a non-Gaussian gate), like we mentioned above. This is simplest in the case of time-independent Hamiltonians. The second way is through the so-called analogue quantum simulation, meaning that one looks for Hamiltonians of systems that naturally have the form one wants, up to the tuning of parameters in the Hamiltonian. Here time-dependence of Hamiltonians is handled automatically. The ease of using this approach depends on the individual experimental platform. For both of these cases, there are many possible physical platforms. See \cite{hangleiter2022analogue} for good discussions on the framework of analogue quantum simulation.\\

Quantum mechanical systems are naturally continuous-variable. Examples of candidates for simulation of qumodes include superconducting circuits \cite{krantz2019quantum}, optical systems \cite{pfister2019continuous, fukui2022building}, optomechanical oscillators \cite{aspelmeyer2014cavity}, atoms in optical lattices \cite{hammerer2010quantum} and vibrational modes of trapped ions \cite{ortiz2017continuous}. These can also be used for simulation of hybrid continuous-variable discrete-variable systems \cite{andersen2015hybrid}. These are systems consisting of both qumodes and qubits, with coupling between qumode and qubit degrees of freedom. We will see these hybrid Hamiltonians for the simulation of inhomogeneous PDEs,  PDEs with derivative in time of more than first-order, and a finite system of PDEs.

\section{Simple examples: Linear convection and higher-order PDEs} \label{sec:convection}

We illustrate first with the linear convection equation -- where a more general approach is not necessary -- to show how simply the continuous-variable framework can be employed. \\

Consider the solutions $u(x_1,...,x_D,t)$ to a $D$-dimensional initial value problem of the linear convection equation
\begin{align}
    \frac{\partial u}{\partial t}+\sum_{j=1}^D a_j \frac{\partial u}{\partial x_j}=0, \qquad u(0,x)=u_0(x), \qquad x_j\in \mathbb{R}^D, \qquad a_j \in \mathbb{R}
\end{align}
with constant $a_j$. We can define an infinite-dimensional vector $\vect{u}$, whose entries are $u(x, t)$ for different $x=(x_1,...,x_D)$ for a specified $t$, so we write
\begin{align}
    \vect{u} \equiv \int u(x_1,...,x_D,t)|x_1, x_2,...,x_D\rangle dx_1...dx_D
\end{align}
with $l_2$ norm $\|\vect{u}(t)\|$. The integral sign $\int=\int_{-\infty}^{\infty}$ unless otherwise specified. By simulating the solutions (in a quantum way), we mean to prepare the normalised quantum state with $D$ continuous-variable modes
\begin{align}
    |u(t)\rangle=\frac{1}{\|\vect{u}(t)\|}\int u(x_1,...,x_D,t)|x_1, x_2,...,x_D\rangle dx_1...dx_D=\frac{\vect{u}(t)}{\|\vect{u}(t)\|}.
\end{align}

We use the observation that $i \hat{p}_j \leftrightarrow\partial/\partial x_j$ here (see Section~\ref{sec:cvintro}) and rewrite the $D$-dimensional convection equation as 
\begin{align} \label{eq:convectionode}
    \frac{d \vect{u}}{d t}=-i \mathbf{H}\vect{u}, \qquad \mathbf{H}=\sum_{j=1}^D \mathbf{H}_j, \qquad \mathbf{H}_j=a_j \hat{p}_j, \qquad \vect{H}_j=\vect{H}^{\dagger}_j.
\end{align}
We note that this can be considered now as an (infinitely large) ODE system since we have now converted differentiation with respect to $x$ by an operator applied to $\vect{u}$. Since $a_j$ is a constant and $\vect{H}$ is time independent, this equation has the solution
\begin{align}
    \vect{u}(t)=e^{-i\vect{H}t}\vect{u}(0), \qquad \text{similarly} \qquad |u(t)\rangle=e^{-i\vect{H}t}|u(0)\rangle
\end{align}
where the latter equality holds since $\|\vect{u}(t)\|=\|\vect{u}(0)\|$ because the convection equation is a conservative equation. This means that preparing $|u(t)\rangle$ only requires a sequence of displacement operators $U=\prod_{j=1}^D U_j $ where $U_j=\exp(-i\mathbf{H}_jt)$. This only requires $D$ single-mode unitary operations $U=\exp(-i a_j \hat{p}_jt)$ on mode $j$ and identity operations on every other mode. \\

That this transformation works is not surprising, since the displacement operator essentially performs a quantum Fourier transform, which is exactly what turns a convection equation into a Schr\"odinger type equation. From this decomposition, it looks like the solution associated with each dimension can evolve separately, without interfering with the behaviour in other dimensions. Thus no entangling gates are necessary. That only single-mode operations are necessary, without mode coupling, is not surprising, since when $a_j$ is a constant, evolution in each dimension is not coupled with other dimensions. Therefore, the equation for each dimension can be solved separately, i.e. $u(t, x_1,...,x_D)=u(t, x_1-a_1t,...,x_D-a_Dt)$. \\

As a example, suppose we want to prepare $|u(t)\rangle$ for $D=2$, given initial conditions $u(0,x)$. Suppose $u(0,x_1, x_2) \propto \exp(-x_1^2/\sigma_1^2-x_2^2/\sigma_2^2)$. Then we can prepare $|u(0)\rangle$ by just preparing a product state of two Gaussian states $|u(0)\rangle \propto \int dx_1 \exp(-x_1^2/\sigma_1^2)|x_1\rangle \int dx_2 \exp(-x_2^2/\sigma_2^2)|x_2\rangle$. To prepare the evolution, one just needs to apply the displacement operator $\exp(-ic_1 \hat{p}_1t)$ onto the first mode and $\exp(-ic_2 \hat{p}_2t)$ onto the second mode of the initial state $|u(0)\rangle$. The resulting state is thus $|u(t)\rangle$. This is simply a product state of the independent evolution of the two initial modes. Note that this simplicity comes from the fact that the initial state was also in a product state. \\

One sees that this linear convection equation is particularly simple for two reasons:

\begin{enumerate}
    \item Behaviour in each dimension is independent of each other, meaning that one can deduce the final $D$-dimensional solution by propagating the time-dependence of each dimension independently. As an example, if the initial data is factorisable (e.g., can use the separation of variables), then the final state is also factorisable. In this case, the solution in $D$ dimensions can be obtained by solving for $D$ differential equations independently, where each equation only depends on one of the original $D$ variables.
    
    \item  $\vect{u}$ evolves via a unitary operation $\exp(-i \vect{H}t)$, which means a quantum simulation can be used directly. This is not true for most PDEs. 
\end{enumerate}

Next we give examples of PDEs which are simple in the sense of point (2) above, where the solution $\vect{u}$ directly evolves through a unitary operation. However, they are not necessarily simple in the sense of point (1).  In the latter case, we can have $x$-dependence in the coefficients and couplings between the modes corresponding to different dimensions are necessary. We will see that this coupling can be achieved with quantum entangling operations. \\

Consider the following linear PDEs with odd powers of derivatives in $x$ and we can obtain a dynamical equation for $\vect{u}$ of the form in Eq.~\eqref{eq:convectionode}. For instance, consider the conservative $D$-dimensional PDEs of the form
\begin{align}
    \frac{\partial u}{\partial t}+\sum_{k=1}^K \sum_{j=1}^D a_{k,j} (x_1,...,x_{j-1},x_{j+1},...,x_D) \frac{\partial^{2k-1} u}{\partial x_j^{2k-1}}=0, \qquad u(0,x)=u_0(x), \qquad x_j \in \mathbb{R}^D, \qquad a_j \in \mathbb{R}
\end{align}
with highest derivative order $2K-1$. The coefficients $a_{k,j}$ can be any $x$-dependent function, except without dependence on $x_j$. Note this equation preserves the $l_2$ norm of $u$, a necessary condition for direct quantum simulation.  This is not necessarily the most physically-motivated variable coefficient PDE example, but we are just using it to illustrate that it is possible for a system to be simple in the sense of point (2) but not point (1). Using the same $i\hat{p}_j \leftrightarrow \partial/\partial x_j$ correspondence, one can rewrite this $D$-dimensional equation as a system similar to Eq.~\eqref{eq:convectionode}
\begin{align}
    & \frac{d \vect{u}}{d t}=-i\vect{H}\vect{u}, \qquad \vect{H}=\sum_{k=1}^K (-1)^{k+1}\sum_{j=1}^D\vect{H}_{kj}=\vect{H}^{\dagger}, \nonumber \\
    &\vect{H}_{kj}=a_{k,j}(\hat{x}_1,...,\hat{x}_{j-1},\hat{x}_{j+1},...,\hat{x}_D) \hat{p}_j^{2k-1}.
\end{align}
As an example, if $a_{2,1}=x_2x_5$, then the corresponding operator $\vect{H}_{21}=\hat{p}_1^{3} \hat{x}_2\hat{x}_5=\vect{H}_{21}^{\dagger}$. Here we note that entangling gates are necessary, so it is no longer simple in the sense of point (1).\\

Although it is not simple in the sense of point (1), this setup in simple in the sense of point (2) due to the hermiticity of $\vect{H}$. The hermitian property comes from the fact that each quadrature is hermitian and they each act on a different mode. If $a_{k,j}$ would depend on $\hat{x}_j$, then $a_{k,j}\hat{p}_{j}^{2k-1}$ is no longer hermitian.  Thus for more general PDEs, we do not have the same equation for $\vect{u}$ in Eq.~\eqref{eq:convectionode}, which is already of a form directly simulable through a Schr\"odinger-like equation, with dynamics determined by a unitary operator $\exp(-i\vect{H}t)$ when $\vect{H}$ is time independent. \\

For more general linear PDEs, we can instead transform them into a Schr\"odinger form through adding a single ancilla qumode. This is a simple transformation called \textit{Schr\"odingerisation}, recently introduced in \cite{jin2212quantum, jin2022quantumtechnical} and analysed for qubit-based systems. We will study this procedure in the following Section~\ref{sec:schrodingerisationreview} -- for PDEs which are generally not simple with respect to either point (1) or point (2) -- and see how it can be naturally used in the continuous-variable setting. In Section~\ref{sec:extended}, we also have an extended Schr\"odingerisation formalism when $D$ ancilla modes are added instead.

\section{Method for general linear PDEs: Schr\"odingerisation for continuous-variable quantum systems} 
\label{sec:schrodingerisationreview}

 A linear homogeneous PDE for $u(x,t)$ with first derivative in $t$ (and any order of derivative in $x$) can be written as
\begin{align} \label{eq:ulinearhomo}
    \frac{\partial u}{\partial t}+\sum_{k=1}^K\sum_{j=1}^D a_{k,j}(x_1,...,x_D) \frac{\partial^k u}{\partial x_j^k}+b(x_1,...,x_D)u=0,
\end{align}
where the coefficients $a_{k,j}<0$ when $k$ is an even integer, for stability of the PDE. Since it will not affect the demonstration of our method, we will not label the sign unless when relevant. We remark that the above coefficients can also depend on time $t$ and our formalism below works in the same way.  \\

By embedding $u$ in a continuous-variable state $\vect{u}$, where $\vect{u}(t)= \int_{-\infty}^{\infty} u(t,x)|x\rangle dx$ and we use $x=x_1,...,x_D$, one obtains  the following equation for $\vect{u}$
\begin{align} \label{eq:uode}
    \frac{d \vect{u}}{dt}=-i\vect{A}(\hat{x}_1,...,\hat{x}_D,\hat{p}_1,...,\hat{p}_D) \vect{u}, \qquad \vect{u}(0)
\end{align}
where $\vect{A}(\hat{x}_1,...,\hat{x}_D, \hat{p}_1,...,\hat{p}_D)$, acting on $\vect{u}$,  is a linear operator consisting of factors of $\hat{x}, \hat{p}$ depending on the form of the original PDE. This is easy to see, since any $\partial^k/\partial x_j^k$ gives a contribution of $(i \hat{p}_j)^k$, and any $x$-dependent coefficient $a(x_1,...,x_D)$ gives a contribution $a(\hat{x}_1,...,\hat{x}_D)$. Thus, for Eq.~\eqref{eq:ulinearhomo}, one has 
\begin{align} \label{eq:Ageneral}
    \vect{A}=\sum_{k=1}^K \sum_{j=1}^D a_{k,j}(\hat{x}_1,...,\hat{x}_D)i^{k+3}\hat{p}_j^k-ib(\hat{x}_1,...,\hat{x}_D)
\end{align}
We will examine $\vect{A}$ explicitly for different classes of PDEs later.\\

\begin{figure}[h] 
\includegraphics[width=14cm]{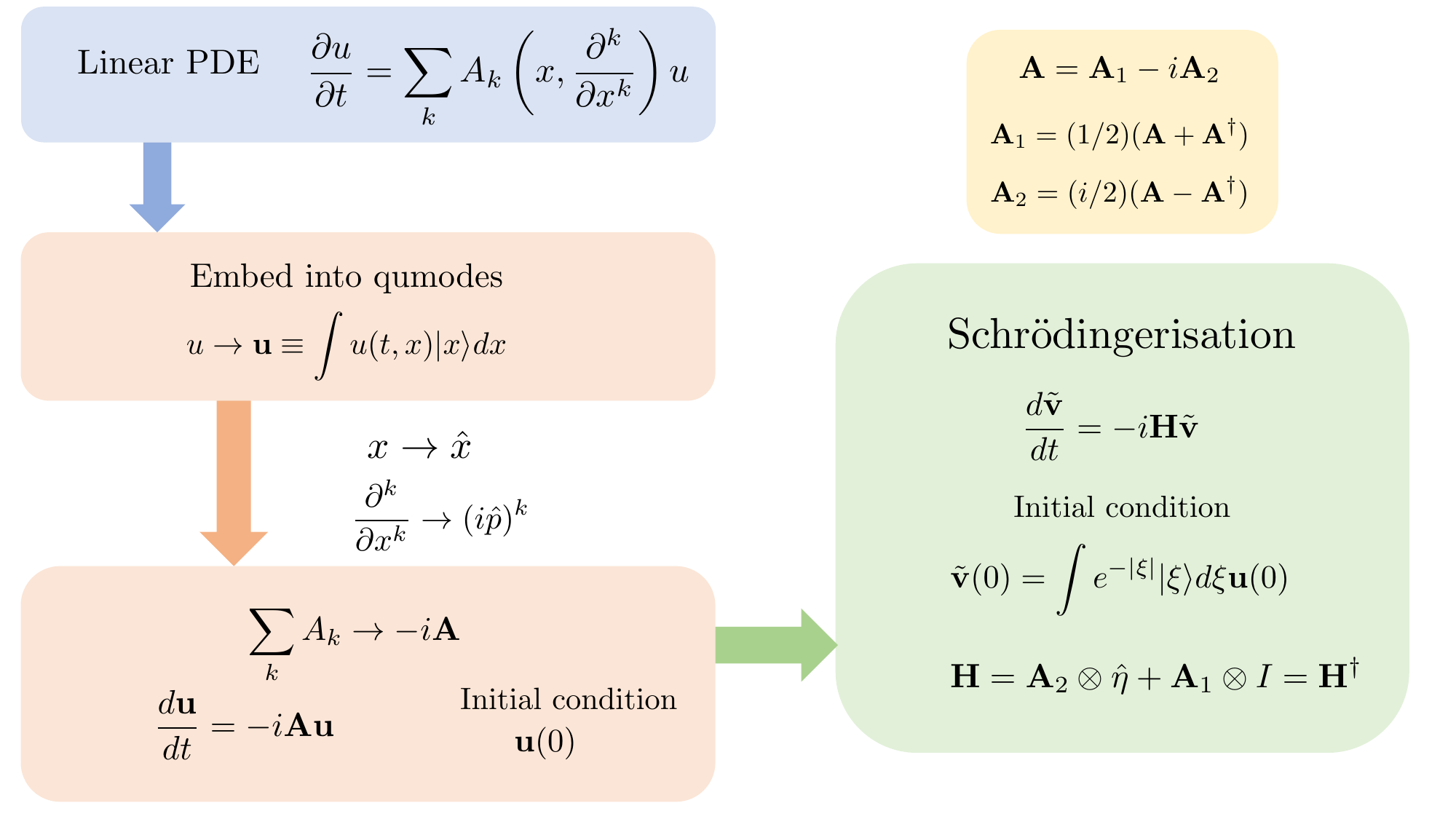} 
\caption{Simple overview of the Schr\"odingerisation procedure} \label{fig:flowchart}
\end{figure}

For the convection equation and certain higher-order PDEs in Section~\ref{sec:convection}, $\vect{A}=\vect{H}$. This means one could directly simulate the evolution of $\vect{u}$ with the unitary operator $\exp(-i\vect{A}t)=\exp(-i\vect{H}t)$. For more general $\vect{A} \neq \vect{A}^{\dagger}$, however, one cannot directly use Hamiltonian simulation. For more general cases, we will use a procedure called \textit{Schr\"odingerisation} \cite{jin2212quantum, jin2022quantumtechnical}. We first observe we can always decompose any operator into its hermitian and antihermitian components 
\begin{align} \label{eq:Adefinition}
    \vect{A}=\vect{A}_1-i\vect{A}_2, \qquad \vect{A}_1=(\vect{A}+\vect{A}^{\dagger})/2=\vect{A}_1^{\dagger}, \qquad \vect{A}_2=i(\vect{A}-\vect{A}^{\dagger})/2=\vect{A}_2^{\dagger}.
\end{align}
We introduce a new parameter $\xi \in \mathbb{R}$ where for $\xi>0$, and define $w(t,x,\xi)=\exp(-\xi)u(t,x)$ and $\vect{w}(t, \xi)=\exp(-\xi)\vect{u}(t)$, which is called a warped phase transformation. Then one can rewrite Eq.~\eqref{eq:uode} as 
\begin{align} \label{eq:vpde}
    \frac{\partial \vect{w}(t, \xi)}{ \partial t}=\vect{A}_2 \frac{\partial \vect{w}(t, \xi)}{\partial \xi}-i\vect{A}_1 \vect{w}(t, \xi).
\end{align}
We  use even extension to the domain  $\xi<0$  (so we can perform the Fourier transform) by extending the initial condition evenly (see \cite{jin2212quantum, jin2022quantumtechnical}) to 
\begin{align}
    \vect{w}(0, \xi)=\vect{w}(t=0, \xi)=\exp(-|\xi|)\vect{u}(t=0)=\exp(-|\xi|)\vect{u}(0).
\end{align}
Let $\tilde{\vect{w}}$ denote the Fourier transform of $\vect{w}$ with respect to $\xi$ with the corresponding Fourier mode $\eta$, i.e. $\tilde{\vect{w}}(t, \eta)=\int d\xi \exp(-i \xi \eta) \vect{w}(t, \xi)$. Then Eq.~\eqref{eq:vpde} becomes the following set of equations for all $\eta$ 
\begin{align}
    \frac{d \tilde{\vect{w}}(t, \eta)}{dt}=-i (\eta \vect{A}_2+\vect{A}_1)\tilde{\vect{w}}(t, \eta), \qquad \tilde{\vect{w}}(0,\eta)=\frac{2}{1+\eta^2}\vect{u}(0).
\end{align}

Now we can define the infinite-dimensional $\tilde{\vect{w}}(t,\eta)$ for all values of $\eta$ by $\tilde{\vect{v}}(t)$, so we can write $\tilde{\vect{v}}(t)= \iint \tilde{w}(t,x,\eta)|x\rangle|\eta\rangle dx d \eta$. Here $\hat{\eta}$ can be any quadrature, e.g. $\cos(\theta)\hat{x}+\sin(\theta)\hat{p}$ for any $\theta \in [0, 2\pi)$. We can define an operator $\hat{\eta}$ such that $\hat{\eta}|\eta\rangle=\eta|\eta\rangle$. This means $(I \otimes \hat{\eta})\tilde{\vect{v}}(t) =\iint \eta \tilde{w}(t,x,\eta)|x\rangle|\eta\rangle dx d \eta$. Then $\tilde{\vect{v}}(t)$ obeys
\begin{align} \label{eq:vtilde0}
   & \frac{d \tilde{\vect{v}}(t)}{dt}=-i (\vect{A}_2 \otimes \hat{\eta}+\vect{A}_1 \otimes I)\tilde{\vect{v}}(t)=-i \vect{H} \tilde{\vect{v}}(t), \qquad \vect{H}=\vect{A}_2\otimes \hat{\eta}+\vect{A}_1 \otimes I=\vect{H}^{\dagger}, \nonumber \\
   & \tilde{\vect{v}}(0)=\int \frac{2}{1+\eta^2}|\eta\rangle d \eta \vect{u}(0)=\int e^{-|\xi|}|\xi\rangle d\xi\vect{u}(0)=|\Xi\rangle\vect{u}(0), \qquad |\Xi\rangle=\int e^{-|\xi|}|\xi\rangle d\xi.
\end{align}
This equation is now of the form where we can directly use Hamiltonian simulation, where $\vect{H}=\vect{H}(\hat{x}_1,...,\hat{x}_D, \hat{p}_1,...,\hat{p}_D)$. In the case the coefficients of the PDE are time-independent, then $\vect{H}$ is also time-independent and the solution is $\tilde{\vect{v}}(t)=\exp(-i\vect{H}t)\tilde{\vect{v}}(0)$. Now $|v(t)\rangle=(1/\|\vect{v}\|) \iint_{-\infty}^{\infty} w(t,x, \xi)|x\rangle |\xi\rangle dx d \xi$ with $\|\vect{v}(t)\|=\|\vect{v}(0)\| \equiv \|\vect{v}\|$. This equality holds since $w(t,x,\xi)$ obeys a conservative equation and hence the $l_2$ norm is preserved. Thus we can write
\begin{align}
    |\tilde{v}(t)\rangle=e^{-i\vect{H}t}|\tilde{v}(0)\rangle, \qquad \vect{H}=\vect{A}_2\otimes \hat{\eta}+\vect{A}_1 \otimes I,  \qquad |\tilde{v}(0)\rangle=|\Xi\rangle  |u(0)\rangle.
\end{align}
Now, for the PDE in Eq.~\eqref{eq:ulinearhomo} for $a_{k,j}, b \in \mathbb{R}$ with $\vect{A}$ given in Eq.~\eqref{eq:Ageneral}, we can easily write the Hamiltonian as
\begin{align}
    \vect{H}=-b\otimes \hat{\eta}+\frac{1}{2}\sum_{k=1}^K\sum_{j=1}^D i^{k}\{a_{k,j}, \hat{p}_j^k \}\otimes \hat{\eta}+\frac{1}{2}\sum_{k=1}^K\sum_{j=1}^D i^{k+1}[\hat{p}_j^k, a_{k,j}]\otimes I.
\end{align}
This form can be greatly simplified as we deal with specific PDEs, as we will see in the following sections. This simple procedure is summarised in Figure~\ref{fig:flowchart}. From this it is also simple to see how, when given any $\vect{H}$, one can work backwards to find the corresponding PDE. \\

Given $|\tilde{v}(t)\rangle$, we can perform an inverse quantum Fourier transform (continuous) on $|\tilde{v}(t)\rangle$ to retrieve $|v(t)\rangle$. This is straightforward to do for instance on optical systems and is essentially just a change in basis from one quadrature to its conjugate. We can take a projection $\hat{P}_{>0}=I \otimes \int_0^{\infty}|\xi'\rangle \langle \xi'| d\xi'$ to select only the $\xi'>0$ parts of the state $|v(t)\rangle$. Here $\xi'$ can represent any quadrature, e.g. momentum. Since  $w(t,x,\xi)=e^{-\xi}u(t,x)$ for $\xi>0$ \cite{jin2022quantum}, then one can see
\begin{align}
   &  \hat{P}_{>0}|v(t)\rangle=\frac{1}{\|\vect{v}\|}\int_{-\infty}^{\infty} \int_0^{\infty}w(t,x, \xi)|x\rangle |\xi\rangle dx d\xi=\frac{1}{\|\vect{v}\|}  \int_0^{\infty} e^{-\xi}|\xi\rangle d\xi \int_{-\infty}^{\infty} u(t,x)|x\rangle dx \nonumber \\
   &=\frac{\|\vect{u}(t)\|}{\sqrt{2}\|\vect{v}\|}|\Xi'\rangle |u(t)\rangle=\frac{\|\vect{u}(t)\|}{\sqrt{2}\|\vect{u}(0)\|}|\Xi'\rangle |u(t)\rangle
\end{align}
where $|\Xi'\rangle=\sqrt{2}\int_{0}^{\infty}\exp(-\xi)|\xi\rangle d\xi$. The second equality uses $\int_0^{\infty}w(t,x,\xi)d\xi=u(t,x)$. Thus the probability of retrieving the state $|\Xi\rangle |u(t)\rangle$ after the projective measurement $\hat{P}_{>0}$ is given by $\left(\|\vect{u}(t)\|/(\sqrt{2}\|\vect{v}\|)\right)^2=(\|\vect{u}(t)\|^2/2)\iint_{-\infty}^{\infty}w^2(0,x,\xi)dx d\xi$. Using the initial condition $w(0,x,\xi)=\exp(-|\xi|)u(0, x)$ for all $\xi$, then one can write $\|\vect{v}\|^2=\int_{-\infty}^{\infty} e^{-2|\xi|} d\xi \int_{-\infty}^{\infty} u^2(0, x)dx=\|\vect{u}(0)\|^2$. 
Thus the probability of retrieving $|u(t)\rangle$ from $|v(t)\rangle$ through $\hat{P}_{>0}$ is $\|\vect{u}(t)\|^2/(2\|\vect{u}(0)\|^2)$. This factor is an inevitable consequence of mapping any PDE into the Schr\"odinger-like conservative PDE. Any method that maps linear PDEs into a Schr\"odinger-like conservative PDE will have this factor. \\

Instead of $\hat{P}_{>0}$, we can also consider a simpler projection $\hat{P}_*=|\xi^*\rangle \langle \xi^*|$ onto a single $\xi=\xi^*>0$, where $\xi^*>0$ is arbitrary. In this case, since $\xi^*>0$, we use $w(t,x,\xi^*)=\exp(-\xi^*)u(t,x)$ and see
\begin{align}
    \hat{P}_*|v(t)\rangle=\frac{1}{\|\vect{v}\|}\int_{-\infty}^{\infty} w(t,x,\xi^*)|x\rangle dx  |\xi^*\rangle=\frac{e^{-\xi^*}} {\|\vect{v}\|} \int_{-\infty}^{\infty}u(t,x)|x\rangle dx=e^{-\xi^*}\frac{\|\vect{u}(t)\|}{\|\vect{u}(0)\|}|u(t)\rangle 
\end{align}
where we used $\|\vect{v}\|=\|\vect{u}(0)\|$ in the last equality. This means that with any measurement outcome $\xi=\xi*>0$ in the ancilla mode, we can retrieve the state $|u(t)\rangle$ with probability $(\exp(-2\xi^*)\|\vect{u}(t)\|/\|\vect{u}(0)\|)^2$. We emphasise that here $\xi^*$ can be any arbitrary $\xi^*>0$, and for any $\xi^*>0$, we can retrieve $|u(t)\rangle$.\\

We can also generalise this to imperfect projective measurements in the $|\xi\rangle$ basis and define $\hat{P}_{\text{imp}}=\int_0^{\infty} f(\xi)|\xi\rangle \langle \xi|$ where $f(\xi)$ models the imperfection in the detector. For instance, it could be a top-hat function or a Gaussian function  with the width or standard deviation denoting the precision of the detector. In this case we still have  $\hat{P}_{\text{imp}}|v(t)\rangle \propto |u(t)\rangle$, where the probability of retrieving $|u(t)\rangle$ is now $(\int_0^{\infty} f(\xi)e^{-\xi} d\xi \|\vect{u}(t)\|/\|\vect{u}(0)\|)^2$. \\

We see that in all these cases, we require the preparation  of the initial state $|\tilde{v}(0)\rangle=|\Xi\rangle|u(0)\rangle$ where $|\Xi\rangle=\int e^{-|\xi|}|\xi\rangle d\xi$. However, $|\Xi\rangle$ is an unusual state which is not necessarily easily attainable in a laboratory setting. However, it is possible to approximate this state for instance by the experimentally accessible Gaussian state $|G\rangle=\int (1/(\sqrt{s}\pi^{1/4}))\exp(-\xi^2/(2s^2))|\xi\rangle d\xi$ with squeezing parameter $s$. Then the quantum fidelity between $|G\rangle$ and $|\Xi\rangle$ is $|\langle \Xi|G\rangle|=\sqrt{2s}\exp(s^2/2)\pi^{1/4}\text{Erfc}(s/\sqrt{2})$. It's clear that there is some optimal $s$ in which we can attain maximum overlap. In fact, the maximum fidelity $|\langle \Xi|G\rangle| \approx 0.986$ is near $1$ at $s \approx 0.925$ which is almost near unity. Now if the initial condition $u(0,x)$ is a Gaussian function with respect to $x$, then $|u(0)\rangle$ can also be efficiently prepared. For a discussion on the error propagation due to using a different ancilla state, see Appendix~\ref{app:initialstatepreparation}.\\

For discussions on error propagation and robustness due to errors in the parameters of Hamiltonian during implementation, see Appendix~\ref{app:robustness}. In this case, we see that the error in the parameter of the Hamiltonian need only scale like $\epsilon \sim O(1/D)$ where $D$ is the dimension of the corresponding PDE to be simulated.\\

We can also generalise the method in this section to deal with inhomogeneous PDEs and higher-order derivatives in $t$ in the following sections.  \\

\begin{figure}[h] \label{fig:homo}
\includegraphics[width=12cm]{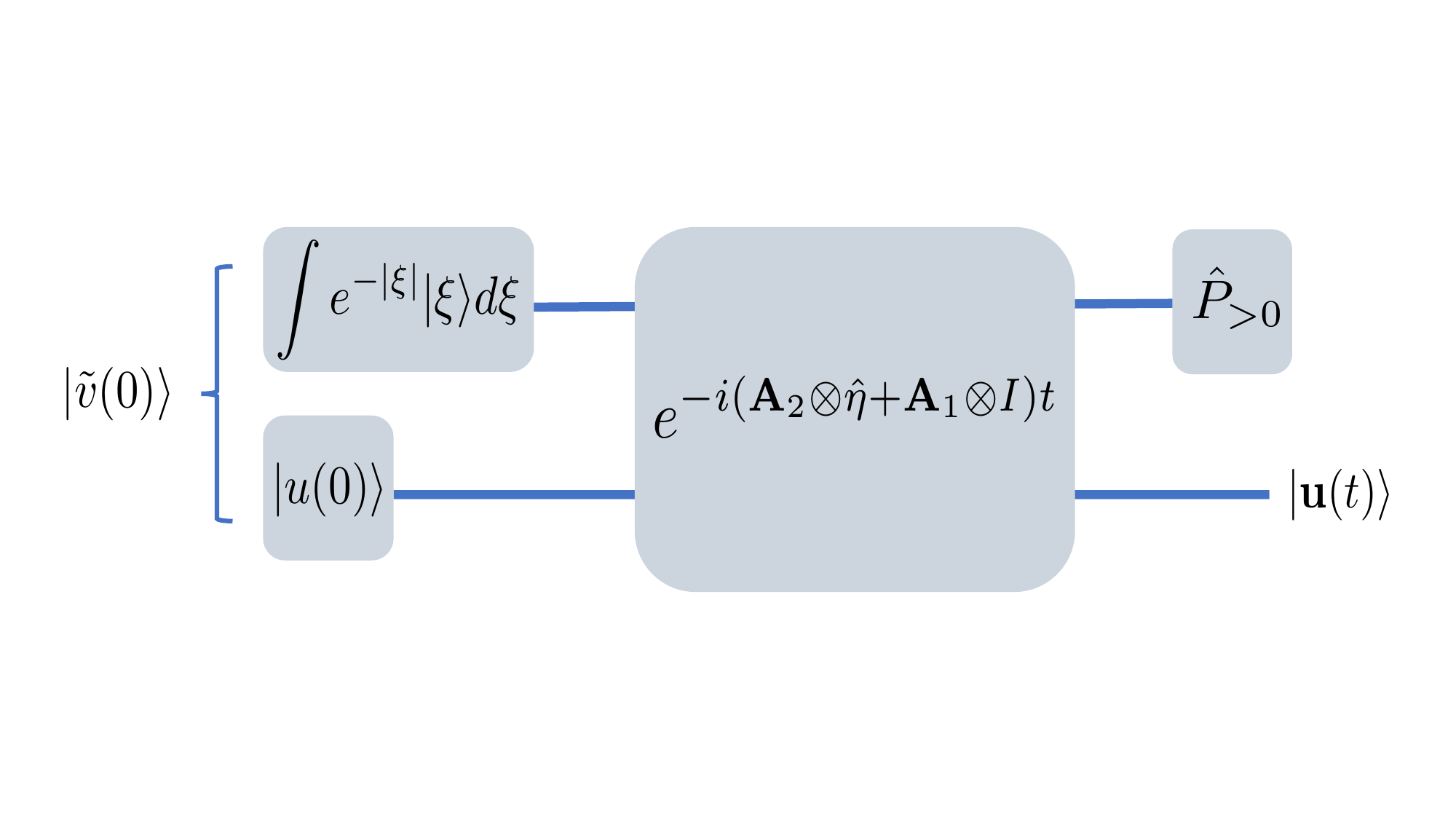}
\caption{Circuit for homogeneous linear PDE with first-order derivatives in time. Here $\hat{P}_{>0}$ requires only projection onto $\xi>0$, and not any particular value of $\xi$.}
\end{figure}

\subsection{Inhomogeneous PDE} \label{sec:inhomo1}

A linear inhomogeneous PDE for $u(x,t)$ with first derivative in $t$  when $f \neq 0$ can be written as
\begin{align} \label{eq:ulinear2}
    \frac{\partial u}{\partial t}+\sum_{k=1}^K\sum_{j=1}^D a_{k,j}(x_1,...,x_D) \frac{\partial^k u}{\partial x_j^k}+b(x_1,...,x_D)u=f(x_1,...,x_D),
\end{align}
where the coefficients $a_{k,j}<0$ when $k$ is an even integer.
We denote that vector giving the value $f(x_1,...,x_D)$ at the point $\vect{x}=(x_1,...,x_D)$ by $\vect{f}$. One can then turn the inhomogeneous equation
\begin{align}
    \frac{d \vect{u}}{dt}=-i\vect{A}(\hat{x}_1,...\hat{x}_D, \hat{p}_1,...,\hat{p}_D)\vect{u}+\vect{f}(\hat{x}_1,...,\hat{x}_D)
\end{align}
into a homogeneous one by absorbing the  inhomogeneous term $\vect{f}$ \cite{jin2022quantumtechnical}. By dilating the system $\vect{u} \rightarrow \vect{y}=\vect{u}\otimes |0\rangle + \vect{f} \otimes |1\rangle$, from Eq.~\eqref{eq:ulinear1} one can  derive
\begin{align}
  &  \frac{d \vect{y}}{d t}= \frac{d}{d t}\begin{pmatrix}
        \vect{u} \\
        \vect{f}
    \end{pmatrix}=-i\vect{B} \vect{y}, \qquad \vect{B}=\begin{pmatrix}
        \vect{A}& iI \\
        0 & 0
    \end{pmatrix}, \qquad \vect{A}=\vect{A}_1-i\vect{A}_2,  \qquad \vect{B}=\vect{B}_1-i\vect{B}_2,  \nonumber \\
    & \vect{B}_1=(\vect{B}+\vect{B}^{\dagger})/2=\vect{B}_1^{\dagger}, \qquad \vect{B}_2=i(\vect{B}^{\dagger}-\vect{B})/2=\vect{B}_2^{\dagger}, \nonumber \\
    & \vect{B}_1=\begin{pmatrix}
        \vect{A}_1 & iI/2 \\
        -iI/2 & 0
    \end{pmatrix}=\vect{A}_1 \otimes \frac{1}{2}(I+\sigma_z)+\frac{I}{2} \otimes \sigma_y, \qquad \vect{B}_2=\begin{pmatrix}
        \vect{A}_2 &  -I/2 \\
         -I/2 & 0
    \end{pmatrix}=\vect{A}_2 \otimes \frac{1}{2}(I+\sigma_z)-\frac{I}{2} \otimes \sigma_x.
\end{align}
We can Schr\"odingerise this new system (homogeneous with respect to $\vect{y}$) where we transform $\vect{y} \rightarrow \vect{v} \rightarrow \tilde{\vect{v}}$ to obtain
\begin{align}
    \frac{d \tilde{\vect{v}}}{d t}=-i\vect{H} \tilde{\vect{v}}, \qquad \vect{H}=\vect{B}_2 \otimes \hat{\eta}+\vect{B}_1 \otimes I=\vect{H}^{\dagger}.
\end{align}
This means that one just needs to include one extra ancilla qubit to the $(D+1)$-qumodes. To recover $|u\rangle$, we first recover $|y\rangle$ using the same method before for homogeneous PDEs and then project onto the $|0\rangle$ qubit. The generating Hamiltonian $\vect{H}$ thus requires an extra coupling to $\sigma_x,\sigma_y, \sigma_z$.  
 
\begin{figure}[h] \label{fig:inhomo}
\includegraphics[width=12cm]{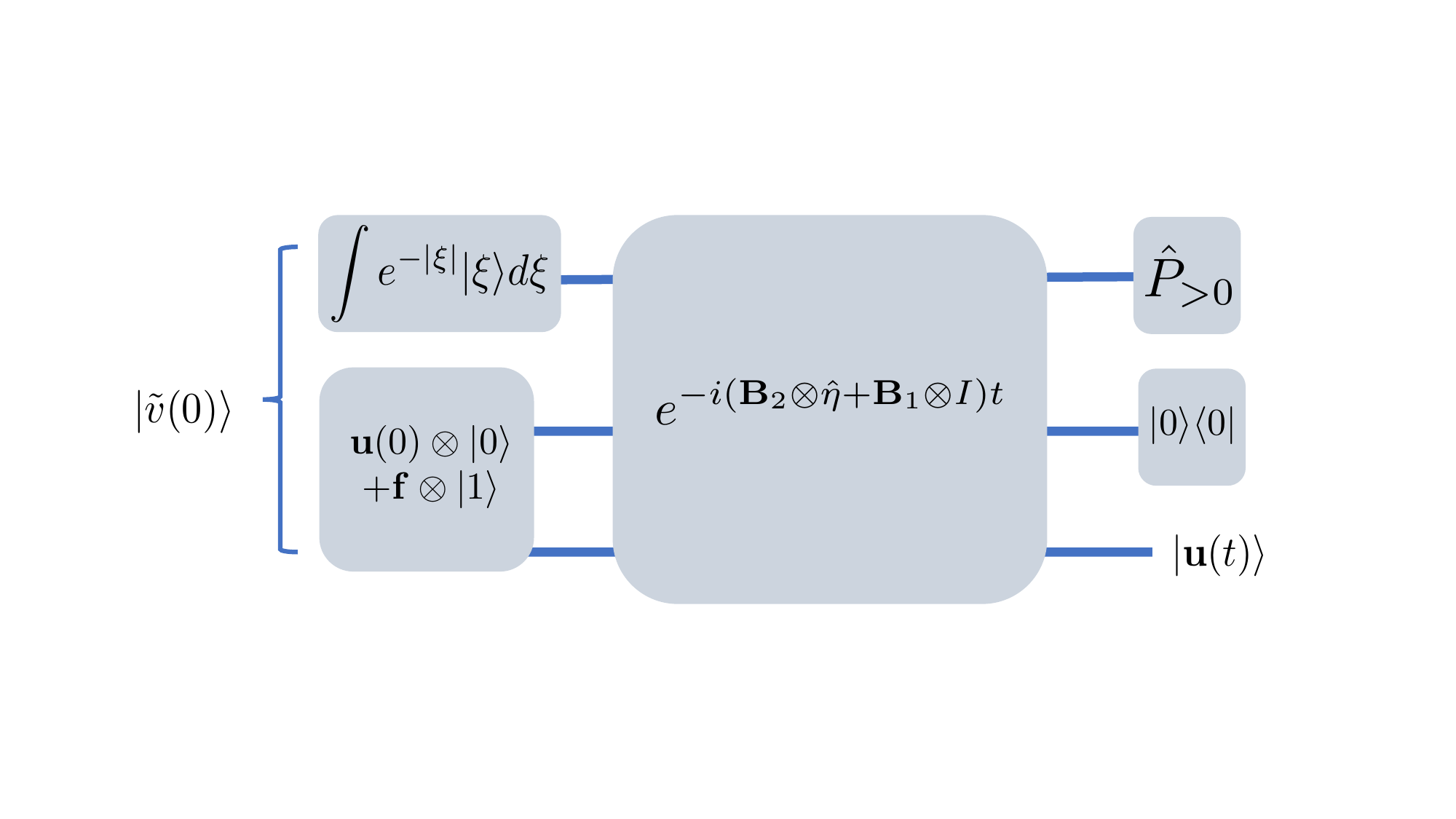}
\caption{Circuit for inhomogeneous linear PDE with first-order derivatives in time. Here $\hat{P}_{>0}$ requires only projection onto $\xi>0$, and not any particular value of $\xi$. In the central register, $|0\rangle \langle 0|$ denotes projection onto the qubit $|0\rangle$.}
\end{figure}

\subsection{PDE with higher-order derivatives in $t$} \label{sec:highert}

We illustrate how our formalism can be applied when the PDE has second-order derivatives in $t$. We look at the following homogeneous linear PDE (includes for example the wave equation as a special case)
\begin{align} \label{eq:t2}
    \frac{\partial^2 u}{\partial t^2}+c_0(x_1,...,x_D) \frac{\partial u}{\partial t}+\sum_{j=1}^d c_j(x_1,...,x_D) \frac{\partial^2 u}{\partial x_j \partial t}+\sum_{l=1}^L\sum_{j=1}^D a_{j,l}(x_1,...,x_D) \frac{\partial^l u}{\partial x_j^l}+b(x_1,...,x_D)u=0,
\end{align}
where the coefficients $a_{j,l}<0$ when $l$ is an even integer.
One can rewrite Eq.~\eqref{eq:t2} into the following equation for $\vect{u}$
\begin{align}
       & \frac{d^2 \vect{u}}{dt^2}+\vect{\Gamma}\frac{d\vect{u}}{dt}+i\vect{A}\vect{u}=0, \qquad \vect{\Gamma} \equiv \vect{c}_0+i\sum_{j=1}^D \vect{c}_j(I^{\otimes j-1}\otimes \hat{p}_j\otimes I^{\otimes D-j}), \qquad \vect{\Gamma}=\vect{\Gamma}_1-i\vect{\Gamma}_2 \nonumber \\
       & \vect{\Gamma}_1=(\vect{\Gamma}+\vect{\Gamma}^{\dagger})/2=\vect{\Gamma}_1^{\dagger}, \vect{\Gamma}_2=i(\vect{\Gamma}-\vect{\Gamma}^{\dagger})/2=\vect{\Gamma}_1^{\dagger}, \nonumber \\
       & \vect{A}=-i\left(\sum_{l=1}^L\sum_{j=1}^D a_{j,l}(\hat{x_1},...,\hat{x}_D)(i\hat{p}_j)^l+b(\hat{x}_1,...,\hat{x}_D)\right)
\end{align}
where $\vect{c}_i=c_i(\hat{x_1},...,\hat{x}_D)$. The goal is to transform this into an equation of the form Eq.~\eqref{eq:uode}.\\

 We can dilate the system $\vect{u} \rightarrow \vect{y}=\vect{u} \otimes |0\rangle+(d\vect{u}/dt) \otimes |1\rangle$. Then  the following equation for $\vect{y}$ can be obtained
    \begin{align}
      &   \frac{d \vect{y}}{d t}=\begin{pmatrix}
          d\vect{u}/dt \\
          d^2\vect{u}/dt^2
      \end{pmatrix}=-i\vect{V} \vect{y}, \qquad \vect{V}=\begin{pmatrix}
            0 & iI \\
            \vect{A} & -i\vect{\Gamma}
        \end{pmatrix}, \qquad \vect{V}=\vect{V}_1-i\vect{V}_2, \qquad \vect{A}=\vect{A}_1-i\vect{A}_2 \\ \nonumber 
        & \vect{V}_1=(\vect{V}+\vect{V}^{\dagger})/2=\vect{V}_1^{\dagger}, \vect{V}_2=i(\vect{V}-\vect{V}^{\dagger})/2=\vect{V}_2^{\dagger}, \qquad \vect{A}_1=(\vect{A}+\vect{A}^{\dagger})/2=\vect{A}_1^{\dagger}, \vect{A}_2=i(\vect{A}-\vect{A}^{\dagger})/2=\vect{A}_2^{\dagger}, \nonumber \\
        & \vect{V}_1=\frac{1}{2}\begin{pmatrix}
            0 & \vect{A}^{\dagger}+iI \\
            \vect{A}-iI & -i(\vect{\Gamma}-\vect{\Gamma}^{\dagger})
        \end{pmatrix}=\vect{A}_1\otimes \sigma_x+\vect{A}_2 \otimes \sigma_y-\frac{I}{2}\otimes \sigma_y+\vect{\Gamma}_2 \otimes \frac{1}{2}(I-\sigma_z), \nonumber \\
        & \vect{V}_2=\frac{1}{2i}\begin{pmatrix}
            0 & -\vect{A}^{\dagger}+iI \\
            \vect{A}+iI & -i(\vect{\Gamma}+\vect{\Gamma}^{\dagger})
        \end{pmatrix}=\vect{A}_2 \otimes \sigma_x-\vect{A}_1 \otimes \sigma_y+\frac{I}{2}\otimes \sigma_x-\vect{\Gamma}_1 \otimes \frac{1}{2}(I-\sigma_z).
    \end{align}
    
    We can then proceed to Schr\"odingerise the system $d \vect{y}/d t=-i\vect{V} \vect{y}$ as before. By applying the transformation $\vect{y} \rightarrow \tilde{\vect{v}}$ in Schr\"oingerisation one has
\begin{align}
     \frac{d \tilde{\vect{v}}}{d t}+i(\vect{V}_2 \otimes \hat{\eta}+\vect{V}_1 \otimes I) \tilde{\vect{v}}=\frac{d \tilde{\vect{v}}}{d t}=-i \vect{H} \tilde{\vect{v}}, \qquad \vect{H}=\vect{V_2}\otimes \hat{\eta}+\vect{V}_1 \otimes I=\vect{H}^{\dagger}.
    \end{align}
It is straightforward to generalise to higher-order derivatives in $t$. 
For instance, it is simple to see that if there is at most an $n^{\text{th}}$-order derivative in $t$, then we need $\log_2(n)$ auxiliary qubits.  

\begin{figure}[h] \label{fig:secondtime}
\includegraphics[width=12cm]{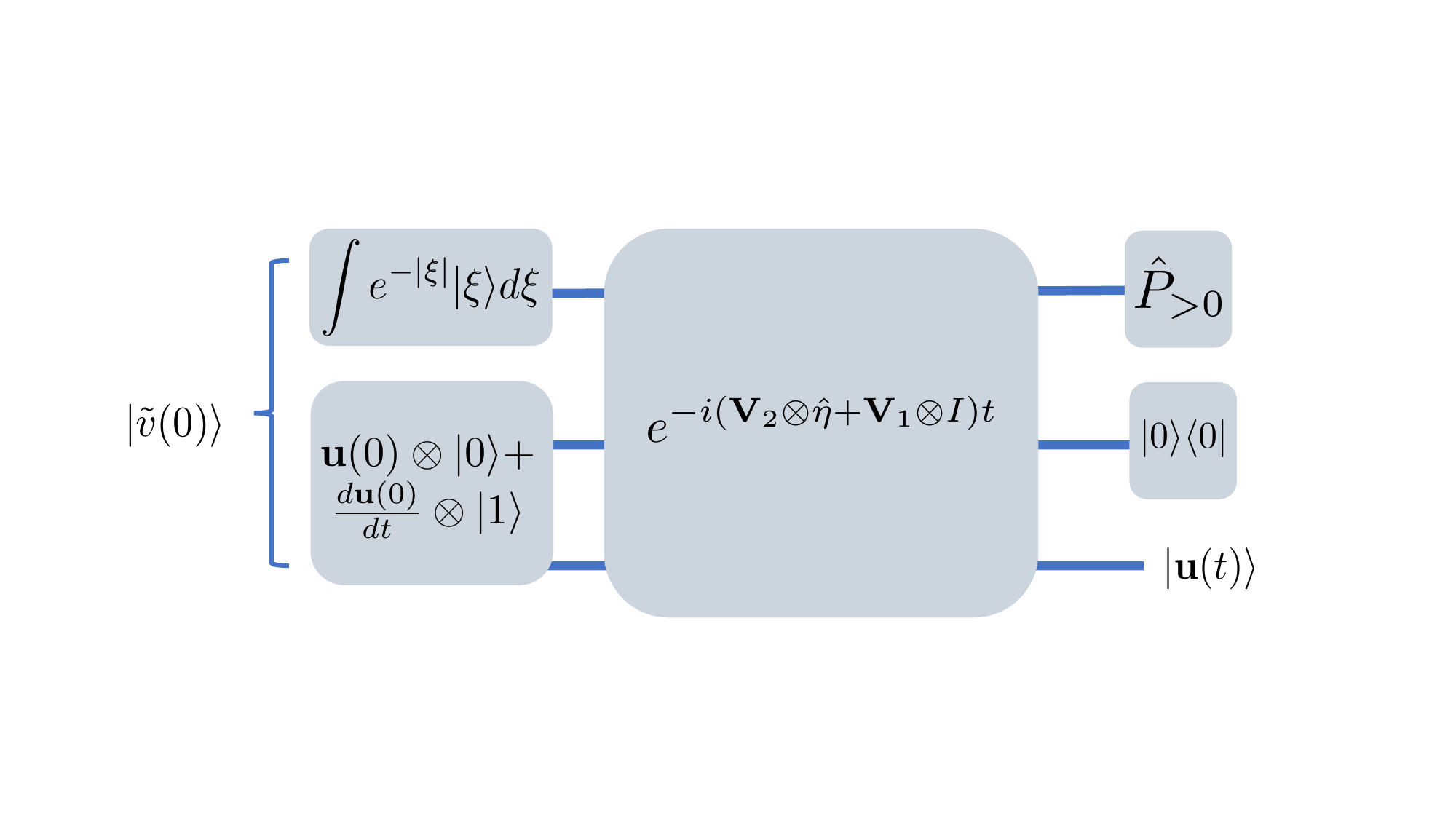}
\caption{Circuit for homogeneous linear PDE with up to second-order derivatives in time. $\hat{P}_{>0}$ requires only projection onto $\xi>0$, and not any particular value of $\xi$. In the central register, $|0\rangle \langle 0|$ denotes projection onto the qubit $|0\rangle$.}
\end{figure}

\section{Examples of Schr\"odingerisation with linear PDEs} \label{sec:example}

In this section, we show examples of different linear PDEs and the Hamiltonians $\vect{H}$ of their corresponding Schr\"odingerised systems. For PDEs with $D$ spatial dimensions, we see that no more than $D+1$ qumodes are necessary. In some cases, only two-mode Gaussian interactions are sufficient, while other PDEs require non-Gaussian interactions. This depends primarily on the $x$-dependence of the coefficients in the PDE and the highest-order spatial derivative in the PDE.

\subsection{Linear first-order PDEs}  \label{sec:generalone}

Here we focus on linear first-order homogeneous PDEs for $u(t, x_1,...,x_D)$
\begin{align} \label{eq:ulinear1}
    \frac{\partial u}{\partial t}+\sum_{j=1}^D a_j(x_1,...,x_D) \frac{\partial u}{\partial x_j}+b(x_1,...,x_D)u=0, \qquad a_j \in \mathbb{R}, b \in \mathbb{C}. 
\end{align}
The inhomogeneous $f \neq 0$ case is treated in Section~\ref{sec:inhomo1} and is straightforward to apply here. Here $a_j$ must be real otherwise the equation is unstable. We can rewrite
\begin{align}
    \frac{d \vect{u}}{d t}=-i\vect{A}\vect{u}, \qquad \vect{A}=\sum_{j=1}^D a_j (\hat{x}_1,...,\hat{x}_D) \hat{p}_j-ib(\hat{x}_1,...,\hat{x}_D), \qquad \vect{A}=\vect{A}_1-i\vect{A}_2
\end{align}
where $\vect{A}_1$ and $\vect{A}_2$ are hermitian operators defined in Eq.~\eqref{eq:Adefinition}. Following Section~\ref{sec:schrodingerisationreview}, we can Schr\"odingerise the system 
\begin{align} \label{eq:vtilde}
    & \frac{d \tilde{\vect{v}}}{dt}=-i\vect{H} \tilde{\vect{v}}, \qquad \vect{H}=\vect{A}_2 \otimes \hat{\eta}+ \vect{A}_1 \otimes I=\vect{H}^{\dagger} \nonumber \\
    &  \vect{A}_1=\frac{1}{2}\sum_{j=1}^D\{\hat{p}_j, a_j(\hat{x}_1,...,\hat{x}_D)\}+\frac{i}{2}(b^*(\hat{x}_1,...,\hat{x}_D)-b(\hat{x}_1,...,\hat{x}_D)), \nonumber \\
    &\vect{A}_2=\frac{i}{2}\sum_{j=1}^D[\hat{p}_j, a_j(\hat{x}_1,...,\hat{x}_D)]+\frac{1}{2}(b(\hat{x}_1,...,\hat{x}_D)+b^*(\hat{x}_1,...,\hat{x}_D))
\end{align}
We can summarise the necessary $\vect{H}$ in Table~\ref{tab:1}, where for brevity we only write out the highest-order terms. In all the examples in the table, there are no parts of the Hamiltonian more difficult than single-mode and two-mode Gaussian operations, except for when $\Re(\gamma_{jk})\neq 0$ in the last line. 

\centering
\begin{center} 
\captionof{table}{Highest-order Hamiltonian terms $\vect{H}$ for simulating linear first-order homogeneous PDEs with mostly Gaussian operations. Below $\hat{\Theta} \equiv (1/2) \sum_{jk} \alpha_{jk}(\hat{x}_j\hat{p}_k+\hat{p}_k\hat{x}_j)\otimes I$, $\alpha_{jk} \in \mathbb{R}$ and $\beta_j$, $\gamma_{jk} \in \mathbb{C}$.}
\label{tab:1}
\begin{tabular}{|c c c|} 
 \hline
  & $\qquad a_j=$ constant  & $\qquad a_j=\sum_{k=1}^D \alpha_{jk} x_k$  \\ [0.5ex] 
 \hline\hline
 $b=$ constant & $\Re(b)I \otimes \hat{\eta}+\sum_j a_j \hat{p}_j \otimes I$ & $\hat{\Theta}$  \\ 
 \hline
 $b=\sum_{j=1}^D \beta_j x_j$ & $\sum_j \Re(\beta_j) \hat{x}_j \otimes \hat{\eta}$ & $\sum_j \Re(\beta_j) \hat{x}_j \otimes \hat{\eta}+\hat{\Theta}$  \\
 \hline
 $b=\sum_{j,k=1}^D \gamma_{jk}x_jx_k$ & \qquad $\sum_{j,k}\Re(\gamma_{jk})\hat{x}_j\hat{x}_k \otimes \hat{\eta}-\sum_{jk} \Im(\gamma_{jk})\hat{x}_j\hat{x}_k\otimes I$ & \qquad $\sum_{jk}\Re(\gamma_{jk})\hat{x}_k \hat{p}_j \otimes \hat{\eta}-\sum_{jk} \Im(\gamma_{jk})\hat{x}_j \hat{x}_k \otimes I+\hat{\Theta}$ \\ [1ex] 
 \hline
\end{tabular}
\end{center}
\justifying Note that if $a$, $b$ are of any higher order polynomials, one would require non-Gaussian gates. For instance, quadratic functions could be made possible with Kerr nonlinearity. For the examples in the table, we see that certain PDEs require only single-mode operations, with generating Hamiltonians like $\hat{\eta}, \hat{x}\hat{p}$, $\hat{x}^2$, while others require pairwise entangling operations like $\hat{\eta} \otimes \hat{x}$. Those that only require single-mode operations are simple in the sense of point (1) in Section~\ref{sec:convection}, like the linear convection equation. \\

A well-known first-order linear PDE for example is the Liouville equation which is Eq.~\eqref{eq:ulinear1} with $b=0$. In the case of linear dependence with $a_j=\sum_{k=1}^D\alpha_{kj}x_k$, $\vect{H}=(1/2)\sum_{j=1}^D \alpha_{jj}I\otimes \hat{\eta}+\hat{\Theta}\otimes I$, which requires pairwise interactions of the form $\hat{x}_j\hat{p}_k+\hat{p}_k\hat{x}_j$ when $j \neq k$. So this is not simple in the sense of point (1) in Section~\ref{sec:convection}, but we require at most Gaussian operations. 

\subsection{Examples of important linear PDEs}
Here we give a few physically important examples of linear PDEs. Most of these have second-order spatial derivatives with the exception of Maxwell's equation.
There do exist physically important PDEs with spatial derivatives of orders higher than two, for example the KdV equation and the quantum hydrodynamic equations (with Bohm potential) which contain third order derivatives, and the Kuramoto-Sivashinsky equation and the Cahn-Hilliard equations which have fourth order derivatives. However, these are typically nonlinear PDEs. We concentrate on linear PDEs in this section with second order spatial derivatives and discuss nonlinear PDEs in Section~\ref{sec:nonlinearsection}.  \\

We will start with problems with first-order derivatives in $t$, including the heat equation, Fokker-Planck equation and Black-Scholes equation. Then we proceed to the wave equation, with second-order in time and space derivatives. For simplicity in these examples we focus on the homogeneous case and the extension to $f\neq 0$ can be done straightforwardly using the formulation in Section~\ref{sec:inhomo1}. Finally, we show Maxwell's equations which is a system of inhomogeneous linear first-order PDEs.

\subsubsection{The heat equation} \label{sec:heat}

Consider the $D$-dimensional general heat equation with a forcing term 
\begin{align}
    \frac{\partial u}{\partial t}-\sum_{i=1}^D\frac{\partial}{\partial x_i}\left(\sum_{j=1}^D D_{ij}(x_1,...,x_D)\frac{\partial u}{\partial x_j}\right)+V(x_1,...,x_D)u=0, \qquad D_{ij}(x_1,...,x_D)>0, \, V(x_1,...,x_D) \in \mathbb{R},
\end{align}
where $D_{ij}$ are the components of the diffusion matrix and we can assume symmetry $D_{ij}=D_{ji}$. This heat equation can be transformed into 
\begin{align}
    & \frac{d \vect{u}}{dt}=-i\vect{A}\vect{u}, \qquad \vect{A}=i\sum_{i,j=1}\hat{p}_i D_{ij}(\hat{x}_1,..., \hat{x}_D)\hat{p}_j-iV(\hat{x}_1,..., \hat{x}_D).
\end{align}
Applying the results of Section~\ref{sec:schrodingerisationreview} for the homogeneous case, the Schr\"odingerised system becomes
\begin{align}
   & \frac{d \tilde{\vect{v}}}{dt}=-i\vect{H}\tilde{\vect{v}}, \qquad \vect{H}=\vect{A}_2 \otimes \hat{\eta}+\vect{A}_1 \otimes I=\vect{H}^{\dagger}, \nonumber \\
   &\vect{A}_1=0 \nonumber \\
   &\vect{A}_2=\sum_{i,j=1}^D \hat{p}_iD_{ij}(\hat{x}_1,...,\hat{x}_D)\hat{p}_j+V(\hat{x}_1,...,\hat{x}_D).
\end{align}
Suppose one has a constant diffusion coefficient $D_{ij}(x)=a \delta_{ij} \in \mathbb{R}$ and a linear driving term $V(x_1,...,x_D)=\sum_{j=1}^Dk_jx_j$, both of which are time-independent. $\vect{A}_1=0$ and we only have the $\vect{A}_2$ term. The Hamiltonian $\vect{H}$ in this case generates an entangling gate where $\vect{H}=\sum_{j=1}^D(a\hat{p}_j^2+k_j\hat{x}_j) \otimes \hat{\eta}$. Thus, the most difficult gate  requires a pairwise third-order operator of the form $\exp(-i \hat{p}^2 \otimes \hat{\eta}t)$. One can either identify an analog quantum system to naturally realise this Hamiltonian, or easily decompose this into more elementary continuous-variable gates. Observe that $\vect{H}=\sum_{j=1}^D\vect{H}_j$ such that $[\vect{H}_i, \vect{H}_j]=0,$ $\forall \, i,j=1,...,D$, so one can factorise $\exp(-i\vect{H}t)=\exp(-i\vect{H}_1t)...\exp(-i\vect{H}_Dt)$. Here $\vect{H}_j=\vect{h}_{j,p}+\vect{h}_{j,x}$, $\vect{h}_{j,p} \equiv a_j \hat{p}_j^2 \otimes \hat{\eta}$, $\vect{h}_{j,x} \equiv k_j \hat{x}_j \otimes \hat{\eta}$. Using the Campbell-Baker-Hausdorff relation with $[\hat{x},\hat{p}^2]=2i\hat{p}$, one obtains the commutation relation $[\vect{h}_{j,x}, \vect{h}_{j,p}]=2ia_jk_j \hat{p}_j \otimes \hat{\eta}=(2i/a_j)\vect{h}_{j,p}$, then  one can decompose $\exp(-i \vect{H}_jt) \propto \exp(i\vect{h}_{j,x}t)\exp(i\vect{h}_{j,p}t/a_j)$. Thus in $D$ dimensions, we can decompose $\exp(-i\vect{H}t)$ into $2D$ elementary continuous-variable gates. These are two-mode entangling gates: $D$ two-mode Gaussian gates of the form $\exp(-i \hat{x} \otimes \hat{\eta}t)$ and $D$ nonlinear gates of the form $\exp(-i\hat{p}^2\otimes \hat{\eta}t)$.\\

\subsubsection{The Fokker-Planck equation} \label{sec:fokkerplanck}

The Fokker-Planck equation gives the time evolution of the probability density function $u(t,x_1,...,x_D)$ of the velocity of a particle under the impact of drag and random forces. In the Fokker-Planck equation below, $\mu_j$ are the components of the drift vector and $D_{j}$ are the components of the diffusion vector
\begin{align}
\frac{\partial u}{\partial t}+\sum_{j=1}^D \frac{\partial}{\partial x_j}(\mu_j (x_1,...,x_D) u)-\sum_{j=1}^D \frac{\partial^2}{\partial x_j^2}(D_{j}(x_1,...,x_D) u)
=0, \qquad \mu_j, D_{j} \in \mathbb{R}.
\end{align}
The drift vector and diffusion coefficients $D_{j}$ only have real-valued components, and $D_j>0$.  Then it is clear 
\begin{align}
& \frac{d\vect{u}}{dt}=-i\vect{A}\vect{u}, \qquad \vect{A}=\sum_{j=1}^D \hat{p}_j \mu_j(\hat{x}_1,...,\hat{x}_D)-i\sum_{j=1}^D \hat{p}^2_j D_{j}(\hat{x}_1,...,\hat{x}_D).
\end{align}
Applying Schr\"odingerisation we see 
\begin{align}
    & \frac{d\tilde{\vect{v}}}{dt}=-i\vect{H}\tilde{\vect{v}}, \qquad \vect{H}=\vect{A}_2\otimes \hat{\eta}+\vect{A}_1\otimes I=\vect{H}^{\dagger}, \nonumber \\
    & \vect{A}_1=\frac{1}{2}\sum_{j=1}^D\{\hat{p}_j,\mu_j(\hat{x}_1,...,\hat{x}_D)\}-\frac{i}{2}\sum_{j=1}^D[\hat{p}^2_j, D_{j}(\hat{x}_1,...,\hat{x}_D)] \nonumber \\
    & \vect{A}_2=\frac{i}{2}\sum_{j=1}^D [\mu_j(\hat{x}_1,...,\hat{x}_D),\hat{p}_j]-\frac{1}{2}\sum_{j=1}^D\{\hat{p}_j^2,  D_{j}(\hat{x}_1,...,\hat{x}_D)\}
\end{align} 
For example, $\mu_j=c_j \hat{x}_j$ with constant $c_j \in \mathbb{R}$ in the case of linear drift. For additive noise $D_{j}=a_j$ where $a_j>0$ is a positive constant and for multiplicative noise, in the case of geometric Brownian motion, $D_j\sim d_j x_j^2$ with real constants $d_j$. In the limit of linear drift and additive noise, $\vect{A}=-\sum_{j=1}^D c_j \hat{p}_j \hat{x}_j-i\sum_{j=1}^D a_j\hat{p}_j^2$ and for Schr\"odingerisation we require the following Hamiltonian acting on $D+1$ modes
\begin{align}
    \vect{H}=\frac{1}{2}\sum_{j=1}^D \left(c_j I-a_j\hat{p}_j^2\right)\otimes \hat{\eta}+\frac{1}{2}\sum_{j=1}^D c_j(\hat{x}_j\hat{p_j}+\hat{p}_j\hat{x}_j)\otimes I.
\end{align}
 When we have linear drift with geometric Brownian motion, the Hamiltonian becomes 
\begin{align}
    \vect{H}=-\frac{1}{2}\sum_{j=1}^D(c_jI-d_j(\hat{x}_j^2\hat{p}_j^2+\hat{p}_j^2\hat{x}^2_j))\otimes \hat{\eta}+\sum_{j=1}^D\left(\frac{c_j}{2}+d_j\right)(\hat{x}_j\hat{p}_j+\hat{p}_j\hat{x}_j)\otimes I
\end{align}
where for both examples, we don't require more than pairwise interactions.

\subsubsection{The Black-Scholes equation} \label{sec:blackscholes}

We can also illustrate our approach with the Black-Scholes equation \cite{black1973pricing}
\begin{align}
    \frac{\partial u}{\partial t}+\frac{1}{2} \sigma^2 x^2 \frac{\partial^2 u}{\partial x^2}+rx \frac{\partial u}{\partial x}-ru=0, \qquad \sigma, r \in \mathbb{R}.
\end{align}
This is a PDE that evaluates the price of a financial derivative, where $r$ and $\sigma$ are constants. Then $\vect{u}$ satisfies
\begin{align}
    \frac{d\vect{u}}{dt}=-i\vect{A}\vect{u}, \qquad \vect{A}=\frac{i}{2}\sigma^2 \hat{x}^2 \hat{p}^2+r\hat{x}\hat{p}-irI.
\end{align}
We can Schr\"odingerise the system
\begin{align}
    & \frac{d \tilde{\vect{v}}}{d t}=-i\vect{H}\tilde{\vect{v}}, \qquad \vect{H}=\vect{A}_2 \otimes \hat{\eta}+\vect{A}_1 \otimes I =\vect{H}^{\dagger} \nonumber \\
    & \vect{A}_1=-\left(\sigma^2+\frac{r}{2}\right)(\hat{x}\hat{p}+\hat{p}\hat{x}), \qquad \vect{A}_2=\frac{1}{4}\sigma^2(\hat{x}^2\hat{p}^2+\hat{p}^2\hat{x}^2)-\frac{r}{2}I.
\end{align}
Since this is a one-dimensional PDE evolving in time, only two qumodes are necessary to simulate $|u\rangle$. The highest-order terms in this Hamiltonian correspond to nonlinear gates of the form $\exp(-i (\hat{x}^2\hat{p}^2+\hat{p}^2\hat{x}^2) \otimes \hat{\eta}t)$. 


\subsubsection{The wave equation} \label{sec:wave}

Consider a $D$-dimensional wave equation of the form
\begin{align}
    \frac{\partial^2 u}{\partial t^2}-\sum_{j=1}^D a_j(x_1,...,x_D) \frac{\partial^2 u}{\partial x_j^2}=-V(x_1,...,x_D)u, \qquad u(0, x), \qquad x_j \in \mathbb{R}^D, a_j \in \mathbb{R}^+, V \in \mathbb{R}. 
\end{align}
This is just a special case of Eq.~\eqref{eq:t2} with $a_{j,2}=a_j$, $b=V$ and all the other coefficients in Eq.~\eqref{eq:t2} are equal to zero. We can then directly use the results in Section~\ref{sec:highert}. Then we see that $\vect{u}$ obeys 
\begin{align}
   & \frac{d^2\vect{u}}{dt^2}=-i\vect{A}\vect{u}, \qquad \vect{A}=-i\left(\sum_{j=1}^D a_j(\hat{x}_1,...,\hat{x}_D)\hat{p}_j^2+V(\hat{x}_1,...,\hat{x}_D)\right) \nonumber \\
   & \vect{A}_1=\frac{i}{2}\sum_{j=1}^D [\hat{p}^2_j,a_j(\hat{x}_1,...,\hat{x}_D)], \qquad \vect{A}_2=\frac{1}{2}\sum_{j=1}^D \{\hat{p}_j^2,a_j(\hat{x}_1,...,\hat{x}_D)\}+V(\hat{x}_1,...,\hat{x}_D).
\end{align}
Following Section~\ref{sec:highert} we see that Schr\"odingerisation can be applied to the dilated system $\vect{u} \rightarrow \vect{y}=\vect{u} \otimes |0\rangle+(d\vect{u}/dt)\otimes |1\rangle$, and we find the following system of $D+1$-qumodes and one qubit
\begin{align}
   &  \frac{d\tilde{\vect{v}}}{dt}=-i\vect{H}\tilde{\vect{v}}, \qquad \vect{H}=\vect{V}_2 \otimes \hat{\eta}+\vect{V}_1 \otimes I,\\
   & \vect{V}_1=\vect{A}_1 \otimes \sigma_x+\left(\vect{A}_2-\frac{I}{2}\right)\otimes \sigma_y, \qquad \vect{V}_2=-\vect{A}_1 \otimes \sigma_y+\left(\vect{A}_2+\frac{I}{2}\right)\otimes \sigma_x.
\end{align}

 The scenarios where $V \neq 0$ could include equations like the Klein-Gordon equation where we use $V=m^2I$, which is easily achieved by including this into the $\vect{A}_2$ term. Then when $a_j$ is a constant $\vect{A}_1=0$ and the highest-order term in $\vect{H}$ would be of the form $\hat{p}^2 \otimes \hat{\eta} \otimes \hat{\sigma}_{x,y}$, which is a coupling between two qumodes and one qubit.

\subsubsection{Maxwell's equations} \label{sec:maxwell}

We can also consider not just a scalar PDE, but rather a system of linear PDEs like Maxwell's equations. Assume a linear medium
with permittivity $\varepsilon$ and permeability $\mu$. The speed of the electromagnetic wave in this medium is then $v=1/\sqrt{\mu \varepsilon}$. We define in Eq.~\eqref{eq:maxwelldefs} the quantities $\mathcal{F}$, $\mathcal{J}$ and $\mathcal{D}$, where $E_{x,y,z}$ and $B_{x,y,z}$ are the $x,y,z$ components of the electric and magnetic fields respectively and $J_{x,y,z}$ are the $x,y,z$ components of the current density and $\rho$ is the charge density. The vectors $\vect{F}$ in bold like $\vect{E}_{x,y,z}$, $\vect{B}_{x,y,z}$, $\vect{J}_{x,y,z}$ and $\vect{\rho}$ represents $\vect{F}=\iiint F(x,y,z)|x\rangle|y\rangle |z\rangle dx dy dz$. Then 
	\begin{align} \label{eq:maxwelldefs}
	& 	\mathcal{F}=\frac{1}{\sqrt{2}}\begin{pmatrix}
			\sqrt{\varepsilon} E_x\\
			\sqrt{\varepsilon} E_y\\
			\sqrt{\varepsilon} E_z\\
			0\\
			(\sqrt{\mu})^{-1} B_x\\
			(\sqrt{\mu})^{-1} B_y\\
			(\sqrt{\mu})^{-1} B_z\\
			0
		\end{pmatrix}, \qquad  \vect{\mathcal{F}}=\sqrt{\varepsilon} \vect{E}_x |0\rangle+\sqrt{\varepsilon} \vect{E}_y |1\rangle+\sqrt{\varepsilon} \vect{E}_z |2\rangle+(\sqrt{\mu})^{-1} \vect{B}_x|4\rangle+(\sqrt{\mu})^{-1} \vect{B}_y|5\rangle+(\sqrt{\mu})^{-1} \vect{B}_z|6\rangle, \nonumber \\
		&\mathcal{J}= \frac{1}{\sqrt{2\varepsilon}}
		\begin{pmatrix}
			J_x\\
			J_y\\
			J_z \\
			0 \\
			0\\
			0\\
			0\\
			-v\rho
		\end{pmatrix}, \qquad \vect{\mathcal{J}}=\vect{J}_x|0\rangle+\vect{J}_y|1\rangle+\vect{J}_z|2\rangle-\vect{v}\vect{\rho}|8\rangle, \nonumber \\
	&	\mathcal{D} = \begin{pmatrix}
			0 &-\partial_z &\partial_y &-\partial_x\\
			\partial_z &0 &-\partial_x &-\partial_y\\
			-\partial_y &\partial_x &0 &-\partial_z\\
			\partial_x &\partial_y &\partial_z &0
		\end{pmatrix}=\begin{pmatrix}
		    0 & i\hat{p}_z & -i\hat{p}_y & i\hat{p}_x \\
      -i\hat{p}_z & 0 & i\hat{p}_x & i\hat{p}_y \\
      i\hat{p}_y & -i\hat{p}_x & 0 & i\hat{p}_z \\
      -i\hat{p}_x & -i\hat{p}_y & -i\hat{p}_z & 0
      \end{pmatrix}=-\hat{p}_z \otimes I \otimes \sigma_y-\hat{p}_x\otimes \sigma_y \otimes \sigma_x+\hat{p}_y \otimes \sigma_y \otimes \sigma_z.
	\end{align}
We consider $\vect{\mathcal{F}}$ as a state of a 3 qumodes and 3 qubits. Then Maxwell's equations can be written as 
	\begin{equation}
		\frac{\partial \mathcal{F}}{\partial t} = v \begin{pmatrix}
			\textbf{0} &\mathcal{D}(1-\bar{\mu})\\
			-\mathcal{D}(1-\bar{\varepsilon}) &\textbf{0}
		\end{pmatrix} \mathcal{F}-\mathcal{J}=v\left(\frac{1}{2}\mathcal{D}(\bar{\varepsilon}-\bar{\mu}) \otimes \sigma_x+\frac{i}{2}\mathcal{D}(2-\bar{\varepsilon}-\bar{\mu})\otimes \sigma_y\right)\mathcal{F}-\mathcal{J}
	\end{equation}
where $\bar{\varepsilon} = \ln \varepsilon/2$, $\bar{\mu} = \ln \mu/2$. Then the following equation governing $\vect{\mathcal{F}}$ with an inhomogeneous term holds:
$\vect{\mathcal{J}}$
\begin{align}
    & \frac{d\vect{\mathcal{F}}}{dt}=-i\vect{A} \vect{\mathcal{F}}-\vect{\mathcal{J}} \nonumber \\
    &\vect{A}=\frac{v(\hat{x}, \hat{y}, \hat{z})}{2} \mathcal{D}(2I-\bar{\varepsilon}(\hat{x}, \hat{y}, \hat{z})-\bar{\mu}(\hat{x}, \hat{y}, \hat{z})) \otimes \sigma_y-i\frac{v(\hat{x}, \hat{y}, \hat{z})}{2}\mathcal{D}(\bar{\varepsilon}(\hat{x}, \hat{y}, \hat{z})-\bar{\mu}(\hat{x}, \hat{y}, \hat{z})) \otimes \sigma_x
\end{align}
One can then apply our formulation to the inhomogeneous case in Section~\ref{sec:inhomo1}. We Schr\"odingerise the system by dilating $\vect{\mathcal{F}} \rightarrow \vect{y}=\vect{\mathcal{F}} \otimes |0\rangle+\vect{\mathcal{J}}\otimes |1\rangle$ and  transform $\vect{y} \rightarrow \tilde{\vect{v}}$ to obtain
\begin{align}
    \frac{d \tilde{\vect{v}}}{dt}=-i \vect{H} \tilde{\vect{v}}, \qquad \vect{H}=\vect{A}_2 \otimes \frac{1}{2}(I+\sigma_z) \otimes \hat{\eta}-\frac{I}{2}\otimes \sigma_x \otimes \hat{\eta}+\vect{A}_1 \otimes \frac{1}{2}(I+\sigma_z) \otimes I+\frac{I}{2}\otimes \sigma_y \otimes I, 
\end{align}
where $\vect{A}_1=(\vect{A}+\vect{A}^{\dagger})/2$, $\vect{A}_2=i(\vect{A}-\vect{A}^{\dagger})/2$. 
This is quantum simulation on a system of 4 qumodes and 4 qubits. \\

For a recent study on Schr\"odingerisation in the qubit formulation and considering physical boundary conditions like perfect conductor and impedance boundaries, as well as considering the interface problem, see \cite{jin2023quantummaxwell}.

\section{PDEs with uncertainty} \label{sec:uqpde}

Often PDEs have coefficients which depend on $L$ stochastic variables, where $L$ might be large. This defines an uncertain PDE, the study of which is the subject of "Uncertainty Quantification" \cite{le2010spectral}. We illustrate our idea in a simple first-order convection equation with uncertain coefficients of the following form
\begin{align} \label{eq:uqpde}
    \frac{\partial u}{\partial t}+\sum_{j=1}^D c_j(z_1,...,z_L, x_1,...,x_D)\frac{\partial u}{\partial x_j}=0, \qquad z_1,...,z_L \sim W(z_1,...,z_L)=\prod_{l=1}^L w(z_l)=\prod_{l=1}^L\frac{\exp(-z_l^2)}{\sqrt{\pi}}, \qquad c_j \in \mathbb{R}
\end{align}
where $z_1,...,z_L$ are random variables each sampled from the Gaussian probability distribution function $w(z_l)$. Typically these equations become very expensive to solve in classical computers as $L$ increases, due to the curse of dimensionality also with respect to $L$. In the Monte-Carlo or other sample based methods such as the stochastic collocation method,  one needs to sample $z_1,...,z_D$ multiple times and solve a new PDE -- due to different $c$ -- for each sample.  \\

Our goal is instead to simulate ensemble-averaged quantities of $u$, without solving Eq.~\eqref{eq:uqpde} multiple times, one for each sample of $z=(z_1,...,z_L)$. We use the polynomial chaos based stochastic Galerkin approach \cite{xiu2002wiener} and call our method the \textit{quantum stochastic Galerkin method}. Let $\vect{n}=(n_1,...,n_L) \in \mathbb{N}_0^L$ be a multi-index. Then any function $u$ (in $l_2$ space with respect to variables $z_1,...,z_D$)  can be expressed as
\begin{align} \label{eq:uexpansion}
    u(t,z_1,...,z_L, x_1,...,x_D)=\sum_{\vect{n}\in \mathbb{N}_0^L} u_{\vect{n}}(t, x_1, \cdots, x_D) P_{\vect{n}} (z_1,...,z_L) .
\end{align}
where 
\begin{align} \label{eq:un}
    u_{\vect{n}}(t,x_1,...,x_D)=\int u(t,z_1,...,z_L,x_1,...,x_D)P_{\vect{n}}(z_1,...,z_L)W(z_1,...,z_L) dz_1...dz_L
\end{align}
with the normalisation
\begin{align} \label{eq:pnormalisation}
    \int P_{\vect{n}}(z_1,...,z_L)P_{\vect{m}}(z_1,...,z_L) W(z_1,...,z_L) dz_1...dz_L=\delta_{\vect{n}\vect{m}}.
\end{align}
$P_{\vect{n}}$ is just the normalized multivariate Hermite polynomials of degree $|\vect{n}|=n_1+\cdots+n_L$. Since $\vect{n} \in \mathbb{N}_0^L$, one can identify $\vect{n}$ to particle numbers, so we can define $\{|\vect{n}\rangle\}_{\vect{n}\in \mathbb{N}_0^L}$ as the Fock or number basis, so $\sum_{\vect{n}\in \mathbb{N}_0^L}|\vect{n}\rangle \langle \vect{n}|=I$. Since $z_1,...,z_D \in \mathbb{R}$, we can associate $z$ with a quadrature value, like position eigenvalues. Thus $\int |z_1,...,z_L\rangle \langle z_1,...,z_L|dz_1...dz_L=I$. This implies $\int \langle \vect{n}|z_1,...,z_L\rangle \langle z_1,...,z_L|\vect{m}\rangle dz_1...dz_L=\delta_{\vect{n}\vect{m}}$. Combining this with the normalisation in Eq.~\eqref{eq:pnormalisation} and $W=\prod_{l=1}^Lw(z_l)$ in Eq.~\eqref{eq:uqpde} allows one to identify
\begin{align} \label{eq:hermitepoly}
     P_{\vect{n}}(z)=\frac{\langle z_1...z_L|\vect{n}\rangle}{W^{1/2}(z_1,...,z_L)}=\prod_{l=1}^LP_{n_l}=\prod_{l=1}^L\frac{\langle z_l|n_l\rangle}{w^{1/2}(z_l)}=\prod_{l=1}^L \frac{H_{n_l}(z_l)}{\sqrt{2^{n_l}n_l!}}, \qquad w(z_l)=\frac{\exp(-z_l^2)}{\sqrt{\pi}},
\end{align}
where $H_{n_l}(z_l)$ are the one-dimensional Hermite polynomials and $P_{n_l}=H_{n_l}(z_l)/\sqrt{2^{n_l}n_l!}$. The last equality is well-known in quantum optics when $|n_l\rangle$ is the Fock state and $|z_l\rangle$ is the position eigenstate \cite{gerry2005introductory}. \\

To recover the ensemble-averaged quantities of $u$, we observe that $P_0(z_l)=H_0(z_l)=1$. From the expansion in Eq.\eqref{eq:un}, one can recover the expectation value of $u$ 
\begin{align}
    u_{\vect{0}}(t, x_1,...,x_D)=\int u(t,z_1,...,z_L, x_1,...,x_D)w(z_1)...w(z_L)dz_1...dz_L \equiv \mathbb{E}_{z_1,...,z_L}(u).
\end{align}
Similarly,  to recover the variance of $u$, using the expansion of $u$ in Eq.~\eqref{eq:uexpansion} and the normalisation condition in Eq.~\eqref{eq:pnormalisation}, it is simple to see that
\begin{align}  \text{Var}_z(u)\equiv \int u^2(t,z_1,...,z_L,x_1,...,x_D)w(z_1)...w(z_l)dz_1...dz_L-\mathbb{E}_{z_1,...,z_L}^2(u)=\sum_{\vect{n}\in \mathbb{N}_0^L, |n|\ge 1} u_{\vect{n}}^2(t,x_1,...,x_D).
\end{align}
This means it is of more physical relevance to solve for $u_{\vect{n}}(t,x_1,...,x_D)$ than the individual $u(t,z_1,...,z_L,x_1,...,x_D)$ values for every random sample of $z$. \\

If $u(t,x_1,...,x_D,z_1,...,z_L)$ solves the PDE in Eq.~\eqref{eq:uqpde}, then using the expansion in Eq.~\eqref{eq:uexpansion} it is simple to show 
\begin{align} \label{eq:zindep}
    & \frac{\partial}{\partial t} u_{\vect{n}}(t,x_1,...,x_D)+\sum_{j=1}^D\sum_{\vect{m} \in \mathbb{N}_0^L}\Lambda_{j\vect{n}\vect{m}}(x_1,...,x_D)\frac{\partial} {\partial x_j}u_{\vect{n}}(t,x_1,...,x_D)=0, \nonumber \\
   & \Lambda_{j\vect{n}\vect{m}}(x_1,...,x_D) =\int c_j(z_1,...,z_L,x_1,...,x_D)w(z_1)...w(z_L)P_{\vect{n}}(z_1,...,z_L)P_{\vect{m}}(z_1,...,z_L)dz_1,...,dz_L
\end{align}
where the coefficients of the PDE are now completely \textit{independent} of $z$! From definition of $\Lambda_{j\vect{n}\vect{m}}$ and Eq.~\eqref{eq:hermitepoly}, one can rewrite
\begin{align} \label{eq:lambdajnm}
    & \Lambda_{j\vect{n}\vect{m}}=\int c_j(z_1,...,z_L,x_1,...,x_D)\langle \vect{m}|z_1,...,z_L\rangle \langle z_1,...,z_L|\vect{n}\rangle dz_1,...,dz_L \nonumber \\
    &=\langle \vect{m}|\int c_j(x_1,...,x_D, \hat{z}_1,...,\hat{z}_L)|z_1,...,z_L\rangle \langle z_1,...,z_L|\vect{n}\rangle dz_1...dz_L=\langle \vect{m}|c_j(x_1,...,x_D, \hat{z}_1,...,\hat{z}_L)|\vect{n}\rangle.
\end{align}
where $\hat{z}_l|z_l\rangle=z_l|z_l\rangle$ is the position operator. \\

Thus instead of solving the $z$-dependent equation multiple times in Eq.~\eqref{eq:uqpde} for every sample of $z_1,...,z_L$, we can solve the $z$-independent equation in Eq.~\eqref{eq:zindep}. We can define $\vect{u}(t)$ as the infinite-dimensional vector in both the Fock and the $x$ bases
\begin{align}
\vect{u}(t) \equiv \int dx \sum_{\vect{n}\in \mathbb{Z}_0^L}u_{\vect{n}}(t,x_1,...,x_D)|\vect{n}\rangle |x_1,...,x_D\rangle
\end{align}
and the corresponding $(d+L)$-qumode quantum state $|u(t)\rangle=(1/\mathcal{N}_u)\vect{u}_{\vect{n}}(t)$ where we assume the normalisation $\mathcal{N}^2_u=\int dx \sum_{\vect{n} \in \mathbb{N}_0^L}|u_{\vect{n}}(t,x_1,...,x_D)|^2 <\infty$.\\

Then it is simple to see from Eq.~\eqref{eq:zindep} and Eq.~\eqref{eq:lambdajnm} that $\vect{u}$ obeys 
\begin{align}
    \frac{d \vect{u}}{dt}=-i\vect{A}\vect{u}, \qquad \vect{A}=\sum_{j=1}^D c_j(\hat{z}_1,...,\hat{z}_L, \hat{x}_1,...,\hat{x}_D)  \hat{p}_j, \qquad \vect{u}(0)=\int \sum_{\vect{n}\in \mathbb{N}_0^L}u_{\vect{n}}(0,x_1,...,x_D)|\vect{n}\rangle |x_1,...,x_D\rangle dx_1...dx_D.
\end{align}
We can easily Schr\"odingerise this equation
\begin{align} \label{eq:uqvtilde}
    &\frac{d\tilde{\vect{v}}}{dt}=-i\vect{H}\tilde{\vect{v}}, \qquad \vect{H}=\vect{A}_2\otimes \hat{\eta}+\vect{A}_1 \otimes I, \qquad \tilde{\vect{v}}(0)=|\Xi\rangle \vect{u}(0), \nonumber \\
    & \vect{A}_1=\frac{1}{2}\sum_{j=1}^D \{c_j(\hat{z}_1,...,\hat{z}_L, \hat{x}_1,...,\hat{x}_D),\hat{p}_j\} \nonumber \\
    & \vect{A}_2=\frac{i}{2}\sum_{j=1}^D [c_j(\hat{z}_1,...,\hat{z}_L, \hat{x}_1,...,\hat{x}_D),\hat{p}_j].
\end{align}
Note that when $c_j$ doesn't depend on $\hat{x}_j$, then  $[c_j,\hat{p}_j]=0=\vect{A}_2$, thus $\vect{A}=\vect{A}^{\dagger}$ and we can directly use Hamiltonian simulation and no Schr\"odingerisation is needed. \\

We emphasise here that although $\vect{A}_1, \vect{A}_2$ depends on the operators $\hat{z}_1,...,\hat{z}_L$, they are only considered as position operators -- \textit{not} random variables -- that act on the number basis $|\vect{n}\rangle$. Eq.~\eqref{eq:uqvtilde} is therefore independent of random coefficients and $c_j(\hat{z}_1,...\hat{z}_L,\hat{x}_1,...,\hat{x}_D)$ is a deterministic operator. However, randomness and information about the distribution $w$ do enter, but \textit{only} through the initial conditions. To prepare the initial state, we first rewrite the initial state in the $|z_1,...,z_L\rangle, |x_1,...,x_D\rangle$ bases
\begin{align}
    \vect{u}(0)=\iint  u(0,z_1,...,z_L,x_1,...,x_D)w^{1/2}(z_1)...w^{1/2}(z_L)|z_1,...,z_L\rangle |x_1,...,x_D\rangle dz_1...dz_Ldx_1...dx_D , \qquad w(z_l)=\frac{e^{-z_l^2}}{\sqrt{\pi}}.
\end{align}
This means that all the randomness of the PDE only resides in the initial state preparation with the inclusion of $w^{1/2}(z_l)$, without needing to solve the PDE in Eq.~\eqref{eq:uqpde} multiple times to estimate ensemble averaged values like the expectation and variance of $u$. For example, if the $x$ and $z$ dependence in the initial state $u(0, z_1,...,z_L, x_1,...,x_D)$ are all Gaussian, then $|u(0)\rangle$ is a Gaussian state since $w$ is also Gaussian. This makes the initial state efficient to prepare. Furthermore, if this initial condition holds, this is the unique initial state to prepare for solving any uncertain linear PDE, and is not dependent on any particular features of the original PDE itself.\\

Another interesting observation is that here it is the  probability amplitude of the distribution of $z_l$, i.e., $w^{1/2}(z_l)$ rather than $w(z_l)$, that is embedded in the amplitude of the pure quantum state $|u(0)\rangle$, which is a uniquely quantum way of embedding the probability distribution. This is a continuous-variable version of the qubit-based Grover-Rudolph embedding \cite{grover2002creating}, which is also considered in quantum walks. A classical sampling of $z_l$, on the other hand, would be equivalent to preparing a density matrix, instead of a pure state, in $z_l$. We leave the discussion of the significance of this embedding for now \cite{quantumstoc2023}.\\

Given the output $\vect{u}(t)$, if one can measure in the number basis $|\vect{n}\rangle\langle \vect{n}|$, one is able to retrieve a state $\propto \int u_{\vect{n}}(t,x_1,...,x_D)|x_1,...,x_D\rangle dx_1...dx_D$, whose amplitudes correspond to already ensemble-averaged physically important quantities. The choice of $\vect{n}=\vect{0}$ for example can give the state whose amplitudes approximate $\mathbb{E}_{z_1,...,z_L}(u)$. \\

  Typically, for smooth, say $C^\infty$, functions $u$, the coefficients in the above expansion in $n_l$ decays exponentially in $n_l$, so only small $n_l$ values need to be kept in the expansion, which can be used to retrieve accurately physically meaningful quantities. Suppose we keep terms in the expansion to $n_l=n_{max}$ for each of the $L$ modes. Then it is known that the error in approximating $u$ with an expansion in $u_{n_l}$ scales like $\sim 1/n_{max}^m$, where $m$ is the regularity of $u$, i.e., all derivatives up to $\partial^m u/\partial z_l^m$ exist and are bounded, but no higher derivatives exist. Thus, to capture most of the ensemble information about $u$, we only need to meausure $n_l$ up to $n_{max}$ which can be small, typically $n_{max} \sim 3-5$ will be enough, for smooth functions $u$.  \\ 

For a simple example, it is known that the following approximation is often used in the Uncertainty Quantification community (either through the Karhunen-Loeve expansion, or the stochastic Galerkin approximation \cite{smith2013uncertainty, xiu2002wiener}) \begin{align}
    c_j(z_1,...,z_L, x_1,...,x_D)\sim \sum_{l=1}^La_{jl}(z_1,...,z_L)b_{jl}(x_1,...,x_D).
\end{align}
Often the Karhunen-Loeve expansion is used and we can assume the linear approximation
\begin{align}
    c_j(z_1,...,z_L, x_1,...,x_D)=c_{j1}(x_1,...,x_D)+\sum_{l=1}^Lc_{j2l}(x_1,...,x_D)z_l, \qquad c_{j1}, c_{j2l} \in \mathbb{R}.
\end{align}
If $c_{j1}(x_1,...,x_D)=\sum_{k=1}^D c_{j1k}x_k$, $c_{j2l}(x_1,...,x_D)=\sum_{k=1}^D c_{j2lk}x_k$ are also linear with constants $c_{j1k}, c_{j2lk}$, this means one can consider Hamiltonians of the general form
\begin{align}
    \vect{H}=\frac{1}{2}\sum_{j,k=1}^D\left(c_{j1k}+\sum_{l=1}^Lc_{j2lk}\hat{z}_l\right)(\hat{x}_k\hat{p}_j+\hat{p}_j\hat{x}_k) \otimes I-\frac{1}{2}\sum_{j=1}^D\left(c_{j1j}+\sum_{l=1}^Lc_{j2lj}\hat{z}_l\right)\otimes \hat{\eta}.
\end{align}

Finally, we remark how continuous-variable quantum systems are naturally suited for this problem through the use of the continuous-variable number and position eigenstates. The inner product of these naturally output  the normalised Hermite polynomials. An analogous quantum algorithm with qubits to approximate the effect of the normalised Hermite polynomials, on the other hand,  would be very complicated. \\

Other classes of linear PDEs and distributions can also be considered following a similar approach and using Schr\"odingerisation. For more details and more general PDEs, see our upcoming 
paper on quantum stochastic Galerkin methods \cite{quantumstoc2023}. 

\section{Nonlinear PDEs} \label{sec:nonlinearsection}

The most natural way to simulate nonlinear PDEs with quantum simulation is to find a way to represent the nonlinear problem in a linear way, since quantum mechanics is fundamentally linear. We must distinguish two types of approaches that converts a nonlinear PDE into a linear PDE. One class of methods necessarily involve approximations (e.g. either through linearisation of the nonlinearity or through discretisation). They include the linear approximation (e.g. Carlemann linearisation) and the linear representation of nonlinear ODEs methods (e.g. through Koopman-von Neumann), where in the latter case the nonlinear ODE is approximated as a finite system of nonlinear ODEs. Typically, these methods are not useful beyond the regime of weak nonlinearity.\\

The second approach is the linear representation for nonlinear PDEs, where no approximations are required to map between the nonlinear and linear PDEs. For our continuous-variable approach, we choose this second method of applying an exact formulation without approximations. \\

While this second approach is the most desirable, it is not simple to find such an exact mapping for all nonlinear PDEs, or know if such mappings even exist. However, we do know some important classes, which include nonlinear Hamilton-Jacobi PDEs (Hamiltonian system) and scalar hyperbolic PDEs (non-Hamiltonian system). Quantum algorithms in the qubit setting and without using Schr\"odingerisation were found for these PDEs in \cite{jin2022quantum, jin2023time}. We now employ our continuous-variable formalism based on Schr\"odingerisation to simulate these nonlinear PDEs.\\

\subsection{The level-set formalism} \label{sec:levelset}

We briefly describe the level set formalism for the nonlinear scalar hyperbolic and nonlinear Hamilton-Jacobi PDEs. The reader can refer to the references \cite{jin2022quantum, jin2023time} for more details on the methodology in the qubit setting. We begin with a $(D+1)$-dimensional scalar nonlinear hyperbolic PDE with $M$ initial conditions 
\begin{align}\label{hyp-PDE2}
  \frac{\partial u^{[k]}}{\partial t} + \sum_{j=1}^D F_j(u^{[k]})\frac{\partial u^{[k]}}{\partial x_j} + Q(x_1,...,x_D,u^{[k]})=0, \qquad  u^{[k]}(0,x)=u^{[k]}_0(x), \quad k=1,...,M
  \end{align}
  We introduce a level set function $\phi(t,x_1,...,x_D,\chi)$, where $\chi \in \mathbb{R}$. It is defined so its zero level set is the solution
  $u$:
  \begin{equation}
        \phi^{[k]}(t,x_1,...,x_D,\chi)=0  \quad {\text{at}} \quad \chi=u^{[k]}(t,x_1,...,x_D)
  \end{equation}
  Then defining $X=(x_1,...,x_D, \chi)$, $a_j=F_j(\chi)$ for $j=1,...,D$ and $a_{D+1}=Q(x_1,...,x_D,\chi)$
  \begin{align}
    \Psi(t, x_1,...,x_D, \chi)=\frac{1}{M}\sum_{k=1}^M \delta(\phi^{[k]}(t,x_1,...,x_D,\chi))
\end{align}
Then $\Psi$ evolves according to the linear PDE in one extra dimension
\begin{align}
     & \frac{\partial \Psi^{[k]}}{\partial t} + \sum_{j=1}^{D+1}  \frac{\partial }{\partial X_j} (a_j(X)\Psi^{[k]})= 0\\
      & \Psi^{[k]}(0,x_1,...,x_D,\chi)=\frac{1}{M}\sum_{k=1}^M \delta(\chi-u^{[k]}_0(x_1,...,x_D)).
\end{align}
If $\vect{\Psi}(t) \equiv \int dX \Psi(t, X)|X\rangle$, then  it satisfies the following linear ODE 
\begin{align} \label{eq:hyperboliclinear1}
    \frac{d \vect{\Psi}}{dt}=-i\vect{A}\vect{\Psi}, \qquad \vect{A}=\sum_{j=1}^{D+1} \hat{P}_ja_j(\hat{X})=\sum_{j=1}^D \hat{p}_jF_j (\hat{\chi})+\hat{\zeta}Q(\hat{x}_1,...,\hat{x}_D, \hat{\chi})
\end{align}
where $i\hat{P}_j=\partial/\partial X_j$, $i\hat{p}_j=\partial/\partial x_j$, $i\hat{\zeta}=\partial/\partial \chi$. Since $\hat{p}_j$ and $\hat{\chi}$ operate on different modes and if $F_j(u^{[k]})$ is a real-valued function, then the first term in $\vect{A}$ is hermitian. Thus when $Q=0$ and $F \in \mathbb{R}$, one has $\vect{A}=\vect{A}^{\dagger}$, so one can directly apply Hamiltonian simulation to Eq.~\eqref{eq:hyperboliclinear1}. Futhermore, one only needs at most pairwise entangling quantum operations of the form $F(\hat{\eta})\hat{p}$. The nonlinearity of the original PDE in Eq.~\eqref{hyp-PDE2} is in the function $F_j$, so we see that due to the term $F_j(\hat{\chi})\hat{p}_j$ in $\vect{H}$, then when $F(u) \propto u$, two-mode Gaussian operations are in fact sufficient. For higher order nonlinearity in $F(u)$, we need to go beyond Gaussian operations to simulate any nonlinearity. \\

However, since $\hat{\chi}$ and $\hat{\zeta}$ operate on the same mode, then the second term depending on $Q$ in $\vect{A}$ is not necessarily hermitian. To deal with general $Q \neq 0$, we can apply Schr\"odingerisation in Section~\ref{sec:schrodingerisationreview} to this system $\vect{\Psi} \rightarrow \tilde{\vect{v}}$ to get 
\begin{align}
   & \frac{d\tilde{\vect{v}}}{dt}=-i\vect{H}\tilde{\vect{v}}, \qquad \vect{H}=\vect{A}_2 \otimes \hat{\eta}+\vect{A}_1 \otimes I=\vect{H}^{\dagger} \nonumber \\
   & \vect{A}_1=\sum_{j=1}^D\hat{p}_jF_j(\hat{\chi})+\frac{1}{2}\sum_{j=1}^D\{\hat{\zeta}, Q(\hat{x}_1,...,\hat{x}_D, \hat{\chi})\}, \qquad \vect{A}_2=\frac{i}{2}\sum_{j=1}^D[ Q(\hat{x}_1,...,\hat{x}_D, \hat{\chi}), \hat{\zeta}].
\end{align} 
Here $\vect{\Psi}$ consists of $D+1$ modes and the Hamiltonian $\vect{H}$ operates on $D+2$ modes. \\

The process for the nonlinear Hamilton-Jacobi equation proceeds similarly. For simplicity, we choose $M=1$. If $S$ obeys the nonlinear Hamilton-Jacobi PDE and $u=\grad S$, then $u$ solves the nonlinear hyperbolic system of conservation equations in gradient form
\begin{align}
    \frac{\partial u_j}{\partial t}+\frac{\partial}{\partial x_j} H(x_1,...,x_D, u_1,...,u_D)=0, \qquad \forall j=1,...,D, \qquad H \in \mathbb{R}
\end{align}
where the function $H=H(x_1,...,x_D, p_1,...,p_2)$ is a Hamiltonian. We now introduce $D$ variables $\chi_1,...,\chi_D \in \mathbb{R}$. 
Then one can define a system of level set functions $\phi=(\phi_1,...,\phi_D)$ where each $\phi_j(t,x_1,...,x_D, \chi_1,...,\chi_D)=0$ at $\chi_j=u_j$. The system $\phi$ solves the linear Liouville equation
\begin{align}
    \frac{\partial \phi}{\partial t}+\sum_{j=1}^D \frac{\partial H}{\partial \chi_j}\frac{\partial \phi}{\partial x_j}-\sum_{j=1}^D\frac{\partial H}{\partial x_j}\frac{\partial \phi}{\partial \chi_j}=0
\end{align}
and the initial data can be chosen as $\phi_j(t=0)=\chi_j-u_j(t=0)$. We can similarly define 
\begin{align}
    \Psi(t,x_1,...,x_D, \chi_1,...,\chi_D)=\prod_{j=1}^D \delta(\phi_j(t,x_1,...,x_D, \chi_1,...,\chi_D)), \qquad \Psi(t=0)=\prod_{j=1}^D \delta(\chi_j-u_j(t=0)).
\end{align}
which obeys the following linear transport PDE
\begin{align}
     \frac{\partial \Psi}{\partial t}+\sum_{j=1}^D \frac{\partial H}{\partial \chi_j}\frac{\partial \Psi}{\partial x_j}-\sum_{j=1}^D\frac{\partial H}{\partial x_j}\frac{\partial \Psi}{\partial \chi_j}=0.
\end{align}
Then 
\begin{align}
    \frac{d \vect{\Psi}}{dt}=-i\vect{H}\vect{\Psi}, \qquad \vect{H}=i\sum_{j=1}^D(\hat{\zeta}_jH(\hat{x}_1,...,\hat{x}_D, \hat{p}_1,...,\hat{p}_D) \hat{p}_j-\hat{p}_jH(\hat{x}_1,...,\hat{x}_D, \hat{p}_1,...,\hat{p}_D)\hat{\zeta}_j)=\vect{H}^{\dagger}.
\end{align}
which acts on $2D$ qumodes. $\vect{H}=\vect{H}^{\dagger}$ since $H$ is interpreted as a Hamiltonian and each of the quadratures is hermitian. Here we can directly use Hamiltonian simulation and no Schr\"odingerisation is actually required. \\

We remark that, since we are solving the Liouville equation, the solution obtained by the level set formalism generates the multivalued solutions \cite{engquist1996multi, sparber2003wigner, jin2003multi, Jin-Osher}, not the viscosity solution 
\cite{lax1973hyperbolic, crandall1983viscosity}, for scalar nonlinear hyperbolic PDEs and Hamilton-Jacobi equations. 

\subsection{Systems of nonlinear ODEs} \label{sec:nonlinearode}

Another way of representing nonlinear PDEs exactly is as a infinite system of nonlinear ODEs. Discretisation methods lead to the approximation of solutions of nonlinear PDEs by a finite system of $N$ nonlinear ODEs. Linearisation methods also lead to solving only a finite number of nonlinear ODEs. Discretisation methods fall largely into two classes: Eulerian and Lagrangian. For example, grid-based Eulerian methods solve the PDEs on a fixed grid while Lagrangian methods are mesh-free methods, like particle methods. Linearisation methods include moment-closure methods and Carlemann linearisation. $N$ can be larger or smaller depending on the discretisation or linearisation technique. \\

This means we first need to know how to simulate a system of $N$ nonlinear ODEs for $\gamma_n(t)$ where each $\gamma_n(t)$ obeys a nonlinear ODE
\begin{align}
    \frac{d\gamma_n(t)}{dt}=F_n(\gamma_0,...,\gamma_{N-1}, t), \qquad n=0,...,N-1
\end{align}
and $F_n$ is a nonlinear function of its arguments. We can use a similar mapping as the level set mapping to solve nonlinear ODEs \cite{jin2022quantum, jin2023time}. This is also related to the Koopman-von Neumann methods, see \cite{jin2023time} for a discussion on the relationship. Here we introduce $N$ auxiliary variables $q_0,...,q_{N-1}$ and a function $\Phi(t, q_0,...,q_{N-1})$ defined by 
\begin{align}
    \Phi(t,q_0,...,q_{N-1})=\prod_{n=0}^{N-1}\delta(q_n-\gamma_n(t)), \qquad q_n \in \mathbb{R}.
\end{align}
Then it is simple to check that $\Phi(t,q_0,...,q_{N-1})$ satisfies, in the weak sense \cite{jin2022quantum}, the \textit{linear} $N+1$-dimensional PDE
\begin{align} \label{eq:nonlinearphi}
    \frac{\partial \Phi(t,q_0,...,q_{N-1})}{\partial t}+\sum_{n=0}^{N-1}\frac{\partial}{\partial q_n}(F_n(q_0,...,q_{N-1})\Phi(t,q_0,...,q_{N-1}))=0.
\end{align}
We can treat this linear PDE with the same methods as before. We can define $\hat{q}_n$ to be a quadrature operator whose eigenstate is $|q_n\rangle$ and its conjugate quadrature operator as $\hat{Q}_n$. Thus $[\hat{q}_n, \hat{Q}_n]=i$ and we can use the correspondence $-i\hat{Q}_n \leftrightarrow \partial/\partial q_n$. Then defining  $\vect{\Phi}(t) \equiv \int \Phi(t,q_0,...,q_{N-1})|q_0,...,q_{N-1}\rangle dq_0...dq_{N-1}$, we see that Eq.~\eqref{eq:nonlinearphi} becomes
\begin{align}
    \frac{d\vect{\Phi}(t)}{dt}=-i\vect{A}\vect{\Phi}(t), \qquad \vect{A}=\sum_{n=0}^{N-1}\hat{Q}_n F_n(\hat{q}_0,...,\hat{q}_{N-1}).
\end{align}
We can easily use the same Schr\"odingerisation methods in Section~\eqref{sec:schrodingerisationreview} $\vect{\Phi} \rightarrow \tilde{\vect{v}}$ and 
\begin{align}
   &  \frac{d \tilde{\vect{v}}}{dt}=-i\vect{H}\tilde{\vect{v}}, \qquad \vect{H}=\vect{A}_2\otimes \hat{\eta}+\vect{A}_1\otimes I=\vect{H}^{\dagger} \nonumber \\
   & \vect{A}_1=\frac{1}{2}\sum_{n=0}^{N-1}\{\hat{Q}_n, F_n(\hat{q}_0,...,\hat{q}_{N-1})\} \nonumber \\
   & \vect{A}_2=\frac{i}{2}\sum_{n=0}^{N-1}[\hat{Q}_n, F_n(\hat{q}_0,...,\hat{q}_{N-1})]
\end{align}
which is a Hamiltonian simulation with $N+1$ qumodes. Given that $F_n$ must be a nonlinear function of its arguments, from the form of $\vect{A}$, it is clear we must necessarily go beyond Gaussian operations, and the order of nonlinearity of the Hamiltonian is either one or two orders larger (depending on if Schr\"odingerisation is needed) than the order of the nonlinearity of the original ODE system.

\begin{table} \label{tab:summary}
\setlength\tabcolsep{0pt} 
\caption{\justifying Summary of resources for simulating PDEs with qumodes, Below $D$ denotes the spatial dimension of the PDE. $K$ is the maximum order of the spatial derivative of the PDE and $L$ denotes the maximum number of stochastic variables in an uncertain PDE. The Hamiltonian $\vect{H}$ refers to the Hamiltonian used in Schr\"odingerisation in the most general scenario. Quantities like $a, b, L, \text{D}, V, \mu, c, Q, H, F$ are $\hat{x}$-dependent operators. They represent the $x$-dependent coefficients of the PDEs. The coefficients $c$ also depends on $\hat{z}$. Here $\hat{\eta}$, $\hat{\zeta}$ can be any quadrature, and $\hat{x}$, $\hat{p}$ can also be any quadrature with the requirement $[\hat{x},\hat{p}]=i$. Below $\{A,B\}\equiv AB+BA$ is the Poisson bracket and $\sigma$ refers to the Pauli matrices. The exact number of terms in $\vect{H}$ in the table below depends on the form of the coefficients of the PDE and must be worked out on an individual basis. If these coefficients can have an expansion with up to $G$ terms in $x$, then there could be $G$ times more terms in $\vect{H}$. \strut}
\begin{tabular*}{\textwidth}{@{\extracolsep{\fill}} l *{5}{C} }
\toprule
Equation
& $Number of qumodes$ 
&  $Max order term in $ \vect{H}
&  $Number of terms in $ \vect{H} \\
\midrule
\textbf{General linear PDEs} 
&

& 
& \\ 
\addlinespace[5mm]
Homogeneous~\ref{sec:schrodingerisationreview}
& D+1
             
& [a_{K,j}, \hat{p}_j^K] \otimes \hat{\eta}, \qquad b \otimes \hat{\eta}
& O(DK)  \\
$1^{\text{st}}$-order time derivative
& 
             
& 
&  \\
\addlinespace
Inhomogeneous~\ref{sec:inhomo1}
& \text{Add 1 more qubit}
& [a_{K,j}, \hat{p}_j^K] \otimes \hat{\eta} \otimes \sigma, \qquad b \otimes \hat{\eta} \otimes \sigma
& O(DK) \\
\addlinespace
$n$-th order time derivative~\ref{sec:highert}
& \quad \text{Add} \log_2(n) \text{ more qubits}
& \text{Similar terms to $1^{\text{st}}$-order} \otimes \sigma^{\otimes \log_2n}
& O(DK) \\
\addlinespace
\addlinespace[5mm]
\textbf{Examples}
& 
& 
&  \\
\addlinespace
Liouville~\ref{sec:generalone}
& D+1
& i[\hat{p},L]\otimes \hat{\eta}, \qquad \{\hat{p},L\}\otimes I
& O(D) \\
\addlinespace[5mm]
Heat~\ref{sec:heat} 
& D+1
& \{\hat{p}\text{D},\hat{p}\}\otimes \hat{\eta}, \qquad V \otimes \hat{\eta}
& O(D+1)  \\
\addlinespace[5mm]
Fokker-Planck~\ref{sec:fokkerplanck}
& D+1
             
& \{\hat{p}^2, \text{D}\}\otimes \hat{\eta}, i[\mu, \hat{p}]\otimes \hat{\eta}, \{\hat{p},\mu\}\otimes I, i[\hat{p}^2, \text{D}]\otimes I
& O(D) \\
\addlinespace
\addlinespace[5mm]
Black-Scholes~\ref{sec:blackscholes}
& 2
& (\hat{x}^2\hat{p}^2+\hat{p}^2\hat{x}^2)\otimes \hat{\eta}& 5  \\
\addlinespace
\addlinespace[5mm]
Wave~\ref{sec:wave} 
& D+1+\text{one qubit}
& \{a, \hat{p}^2\}\otimes \hat{\eta} \otimes \sigma, \quad V\otimes \eta \otimes \sigma, \quad i[a, \hat{p}^2]\otimes I \otimes \sigma
& O(D)  \\
\addlinespace
\addlinespace[5mm]
Maxwell ($D=3$)~\ref{sec:maxwell}
& 4 \text{ qumodes} + 4 \text{ qubits}
& \text{See Section~\ref{sec:maxwell}} 
& 30 \\
\addlinespace
\addlinespace[5mm]
\textbf{Uncertain linear PDE}~\ref{sec:uqpde}\\
\text{Convection}
& D+L+1
& i[c,\hat{p}]\otimes \hat{\eta}, \qquad \{c, \hat{p}\}\otimes I
& O(DL)  \\
\addlinespace
\addlinespace[5mm]
\textbf{Nonlinear}
& 
& 
&  \\
\addlinespace
Scalar hyperbolic~\ref{sec:levelset}
& D+2
& i[Q, \hat{\zeta}]\otimes \hat{\eta}, \qquad \{Q, \hat{\zeta}\}\otimes I
& O(D) \\
\addlinespace
Hamilton-Jacobi~\ref{sec:levelset}
& 2D
& i[\hat{\zeta}H\hat{p}-\hat{p}H\hat{\zeta}]
& O(D)  \\
\addlinespace
$N$ nonlinear ODEs~\ref{sec:nonlinearode}
& N+1
& i[Q,F]\otimes \hat{\eta}, \qquad \{Q, F\}\otimes I
& O(N)  \\
\addlinespace
\bottomrule
\end{tabular*} \label{tab:summary}
\end{table}

\section{Extended Schr\"odingerisation} \label{sec:extended}

In the previous sections using Schr\"odingerisation, we only required a single auxiliary qumode with quadrature operators $\hat{\eta}, \hat{\xi}$. Thus a linear PDE with $D$ spatial dimensions is simulable on a $D+1$-qumode quantum device. This is the smallest number of auxiliary qumodes required, so from the standpoint of complexity for universal analog quantum simulation, this is the most favourable. However, it may not be suitable for direct simulation on some quantum systems for large $D$. This is because the Schr\"odingerisation in Section~\ref{sec:schrodingerisationreview} in general requires $O(D)$ couplings from other modes onto this single mode. For analog quantum systems where multiple couplings to one specific mode is difficult, we can instead modify our Schr\"odingerisation procedure so we have instead $O(D)$ coupling terms with $D$ \textit{distinct} auxiliary modes. We call this the \textit{extended Schr\"odingerisation} procedure. \\

Here, instead of introducing a single $\xi \in \mathbb{R}$, we introduce $D$ auxiliary variables $\xi_1,...,\xi_D \in \mathbb{R}$ and we can define the extended warped phase transformation as $\vect{w}(t, \xi_1,...,\xi_D)=\exp(-\xi_1)...\exp(-\xi_D)\vect{u}(t)$ for $\xi_1,...,\xi_D>0$. We follow a similar derivation as before. From Section~\ref{sec:schrodingerisationreview}, for a general homogeneous linear PDE with first-order time derivative and maximum $K^{\text{th}}$-order spatial derivative, we can write
\begin{align} \label{eq:uodeextended}
    \frac{d\vect{u}}{dt}=-i\vect{A}\vect{u}, \qquad \vect{A}=\sum_{j=1}^D \vect{a}_j, \qquad \vect{a}_j=\sum_{k=1}^K\left(a_{k,j}(\hat{x}_1,...,\hat{x}_D)i^{k+3}\hat{p}^k_j-\frac{i}{K}b(\hat{x}_1,...,\hat{x}_D)\right).
\end{align}
We can decompose each $\vect{a}_j=\vect{a}_{j1}-i\vect{a}_{j2}$ into its hermitian and anti-hermitian components, where $\vect{a}_{j1}=(1/2)(\vect{a}_j+\vect{a}_j^{\dagger})$, $\vect{a}_{j2}=(i/2)(\vect{a}_j-\vect{a}_j^{\dagger})$. Note here $\vect{a}_{j1}$ corresponds to odd-derivative terms, while $\vect{a}_{j2}$ corresponds to the even-derivative terms. We observe that for every $\xi_j$, $j=1,...,D$, we can write $\vect{w}=-\partial \vect{w}/\partial \xi_j$. Thus we can rewrite Eq.~\eqref{eq:uodeextended} as
\begin{align}
    \frac{\partial \vect{w}(t,\xi_1,...,\xi_D)}{\partial t}=\sum_{j=1}^D \vect{a}_{j2} \frac{\partial \vect{w}(t,\xi_1,...,\xi_D)}{\partial \xi_j}-i\vect{a}_{j1}\vect{w}(t,\xi_1,...,\xi_D).
\end{align}
We can similarly perform an even extension to the domain $\xi_1,...,\xi_D<0$ by extending the initial condition evenly
\begin{align}
    \vect{w}(0,\xi_1,...,\xi_D)=\exp(-|\xi_1|)...\exp(-|\xi_D|)\vect{u}(0).
\end{align}
If $\tilde{\vect{w}}$ denotes the Fourier transform of $\vect{w}$ with respect to $\xi_1,...,\xi_D$ with corresponding Fourier modes $\eta_1,...,\eta_D$, then for all $\eta_1,...,\eta_D$
\begin{align}
    \frac{d\tilde{w}(t,\eta_1,...,\eta_D)}{dt}=-i\sum_{j=1}^D(\eta_j\vect{a}_{j2}+\vect{a}_{j1})\tilde{\vect{w}}(t,\eta_1,...,\eta_D), \qquad \tilde{\vect{w}}(0,\eta_1,...,\eta_D)=\left(\frac{2}{1+\eta^2_1}\right)...\left(\frac{2}{1+\eta^2_D}\right)\vect{u}(0).
\end{align}
Now we can also define an extended $\tilde{\vect{v}}(t)\equiv \int \tilde{w}(t,x_1,...,x_D,\eta_1,...,\eta_D)|x_1...x_D\rangle|\eta_1...\eta_D\rangle$. Thus $\tilde{\vect{v}}(t)$ now obeys a Schr\"odinger-like equation
\begin{align}
     \frac{d\tilde{\vect{v}}(t)}{dt}=-i\vect{H}\tilde{\vect{v}}, \qquad \vect{H}=\sum_{j=1}^D \vect{h}_j=\vect{H}^{\dagger}, \qquad \vect{h}_j=\vect{a}_{j2}\otimes \hat{\eta}_j+\vect{a}_{j1}\otimes I, \qquad \tilde{\vect{v}}(0)=|\Xi\rangle^{\otimes D}\vect{u}(0)
\end{align}
which now operates on $2D$ qumodes instead of $D+1$ qumodes. However, the advantage with this approach is that each $\vect{h}_j$ corresponds to coupling to a different auxiliary mode. For example, for those PDEs that require at most pairwise couplings, we only need to simulate $O(D)$ interactions between $D$ completely different pairs of modes instead of $O(D)$ interactions between one particular auxiliary mode and $D$ other modes. The extension to various PDEs is straightforward. If the Hamiltonian of the original Schr\"odingerised system has terms including  $\sum_{j=1}^D F_j(\hat{x})\hat{p}_j^K\otimes \hat{\eta}$, then the corresponding modified terms in the extended Schr\"odingerised scheme would be of the form $\sum_{j=1}^D F_j(\hat{x})\hat{p}_j^K\otimes \hat{\eta}_j$. \\

The projection back onto $|u(t)\rangle$ through $\hat{P}_{>0}=I\otimes \int_0^{\infty} |\xi_1...\xi_D\rangle \langle \xi_1...\xi_D|d\xi_1...d\xi_D$ then works in the same way, except now the probability of success in retrieving $|u(t)\rangle$ becomes $\|\vect{u}(t)\|^2/(2^D\|\vect{u}(0)\|)$. The initial state preparation for $|\Xi\rangle^{\otimes D}$ can still be done using Gaussian states $|G\rangle^{\otimes D}$ for example. However, in this case, the new fidelity becomes $|\Xi|G\rangle|^D$, which decays with $D$. If $|\langle \Xi|G\rangle| \approx 0.986$ is possible at the optimal $s$, then for the preparation of $|\Xi\rangle^{\otimes D}$ for $D=5$, we can reach a fidelity of $\approx 0.932$. Thus, for very high $D$, this extended approach is much more costly and the original Schr\"odingerisation scheme in Section~\ref{sec:schrodingerisationreview} is preferred. However, for some physical systems where the extended Schr\"odingerisation might be easier to physically implement in the near term, low $D$ demonstrations are still possible.

\section{Discussion}
In our formulation, we considered primarily the cost in preparation of the state $|\tilde{v}(t)\rangle$ through Hamiltonian $\vect{H}$ simulation when given the initial state $|\Xi\rangle|u(0)\rangle$. We have also outlined a procedure of retrieving the state $|u(t)\rangle$ whose amplitudes are proportional to the solutions of the PDE. In cases where $\vect{H}$ can be realised directly in analogue quantum simulation, we don't need to consider gate count. The procedure is efficient in the number of qumodes $D+1$. When extended Schr\"odingerisation is used, there are only  $2D$ qumodes. However, when direct quantum simulation is difficult, then a concatenation of continuous-variable quantum gates is required. The exact number needs to computed on an individual basis since it also depends on the form of the coefficients of the PDE. Given that we know the number of terms of the Hamiltonian scales like $O(D)$, and if the terms in the Hamiltonian are pairwise interactions and almost commute, then we expect the gate count to also be of order $O(D)$, where each gate consists of no more than pairwise interactions. A detailed analysis will be left to future work and to see in particular which classes of PDEs are more efficient than others to simulate and which are not efficient. Note that for classical finite difference methods (performed on classical digital devices), the scaling is exponential in $D$. It is also interesting to consider that, in continuous-variable quantum computation, non-Gaussian operations that are necessary for potential quantum speedup \cite{bartlett2002efficient}, otherwise efficient classical simulation can suffice. It is very simple to see in this formulation which part of the PDE requires non-Gaussian gates and which types of initial conditions require non-Gaussian state inputs. This contrasts with qubit-based formulations where it is not clear which aspect of the PDE requires non-Clifford gates, which are the qubit analogues to non-Gaussian gates. We leave discussions on this and its connections to efficient simulability of differential PDEs to future analysis.\\

Independently of the above, our main aim is not in fact any direct comparisons to qubit-based or other classical algorithms based on discretised PDEs. Rather, this is a proposal for a new way to simulate PDEs with analog degrees of freedom that can be achieved beyond using the canonical von Neumann architecture. In the case of Gaussian initial conditions and where only Gaussian operations are required (as is in the case for some linear first-order homogeneous PDEs in Section~\ref{sec:generalone}), our formalism can be considered a novel classical analog simulation technique, which can extend to exploiting quantum characteristics by including non-Gaussian gates with higher-order PDEs. \\

We emphasise that, like in the qubit-based approaches to PDEs, our algorithm is currently also in the form of a quantum subroutine: that we are preparing quantum states who amplitudes are related to the solution of the PDE. The next stage is to consider efficient measurements on these final states that can capture for us observables. Recent methods based on entangled CV measurements for example \cite{wu2023quantum} be used a step forward, which we leave to the next stage of investigation. \\ 

 
We observe that when there is time-dependence in the coefficients of the PDE, our Schr\"odingerisation procedure works in the same way and in this case we find $d \tilde{\vect{v}}(t)/dt=-i\vect{H}(t)\tilde{\vect{v}}$ where now $\vect{H}$ is time-dependent coefficient. When there is an existing physical system with a corresponding time-dependent Hamiltonian $\vect{H}(t)$, then the simulation is straightforward. However, if we require a decomposition of the corresponding unitary into smaller unitaries time-ordered operators are needed and it is not so simple. It is less studied in the continuous-variable scenario and we leave this for future analysis. \\

An important aspect of solving PDEs is the consideration of boundary conditions. Using Schr\"odingerisation to deal with both physical \cite{jin2023quantumphysical} and artificial \cite{jin2023quantumartificial} boundary conditions have been investigated by the authors
in the case of qubits. How this can be extended to analog systems can be left for future work when we study more specific systems. It is also important to remark that if one wishes to simulate analog quantum systems under artificial boundary conditions, then using another analog quantum system via Schr\"odingerisation is necessary. This is because a classical analog (i.e., not digital) system even in low dimensions could have difficulty in efficiently simulating the system along with reproducing uniquely quantum features.\\


 An important consideration for near term realisation is to ask if existing analog quantum systems can be repurposed to simulate linear PDEs, without decomposing the corresponding unitary into smaller gates. Our formalism can be very easily used to identify, presented by a given $\vect{H}$, the corresponding linear PDE, if it exists. This can be done by following the arrows backwards in the procedure in Fig.~\ref{fig:flowchart}. We leave discussions of realistic errors beyond the ideal limit and noise and any possible truncation in the physical degrees of freedom to future work. \\

As a very simple example, suppose we have a quantum optomechanical system with driving $\vect{H}_{OM}=\Delta(\hat{x}_a^2\otimes I+ \hat{p}_a^2 \otimes I)+\Omega(I \otimes \hat{x}_b^2+ I \otimes \hat{p}_b^2)+g_1 \hat{x}_a \otimes \hat{x}_b+g_2(\hat{p}_a \otimes I), \Delta, \Omega, g_1, g_2 \in \mathbb{R}$, 
where modes $a, b$ are the photonic and mechanical modes respectively. Tuning $\Omega \rightarrow 0$, this is able to simulate a Schr\"odinger equation with a complex potential and a convection term
   $ \partial u/\partial t=i\partial^2u/\partial x^2+g_2\partial u/\partial x-(g_1x+i\Delta x^2)u$. Complex potentials can model absorbing boundary conditions and is a technique in artificial boundary conditions and can also be used to model dissipative effects like optical phonon scattering. Coupled superconducting circuits have a similar Hamiltonian and can include more nonlinearities. If only independent pairwise couplings are considered in $\vect{H}$, then the extended Schr\"odingerisation method can be used. \\

   There are also many other platforms to explore and we leave this as a challenge for experimentalists to consider how their own quantum systems -- which only needs a system size of $D+1$ continuous degrees of freedom -- can be repurposed for the simulation of linear PDEs with $D$ spatial dimensions or a system of $D$ nonlinear ODEs. \\

   In addition being applicable to analog degrees of freedom, this Schr\"odingerisation formalism is also flexible enough to work completely in the qubit formulation, as well as having a straightforward hybrid discrete-variable continuous-variable formulation. This makes it simple to adapt to many different platforms. In fact, such hybrid platforms can also be used for other problems like applications to linear algebra \cite{jin2023quantum}. It is also applicable to non-autonomous linear systems, like quantum dynamics with time-dependent Hamiltonians \cite{2023timedependent}.\\
   
   Finally, we remark that, historically, analog devices dominated computation before digital computation, and digital devices didn't completely dominate analog computation until the 1980s, and ideas about quantum computation only emerged after the analog era faded. Today there is a resurgent interest in classical analog proposals as we move beyond the von Neumann architecture, for example see \cite{bournez2021survey}. In quantum computation, it is reasonable to consider perhaps that, just like in classical computation, the first useful devices would also be analog. Since analog and digital models are fundamentally different, it is not sensible to compare them directly -- either classical or quantum algorithms -- but rather it is important to cultivate both kinds of algorithms to grow our toolbox for computational methods. We encourage the reader to consider a more concerted effort toward analog quantum algorithms to enrich methods for scientific computing.

\section*{Acknowledgements}

NL thanks Bill Munro for interesting discussions and suggestions. SJ was partially supported by the NSFC grant No. 12031013, the Shanghai Municipal Science
and Technology Major Project (2021SHZDZX0102), and the Innovation Program of Shanghai Municipal Education Commission (No. 2021-01-07-00-02-E00087). NL acknowledges funding from the Science and Technology Program of Shanghai, China (21JC1402900).
Both authors are also supported by the Fundamental Research Funds for the Central
Universities.

\bibliography{Ref}
 
\appendix

\section{Error estimates for initial state preparation}\label{app:initialstatepreparation}

From section \ref {sec:schrodingerisationreview}, by using $|G\rangle$ instead of $|\Xi\rangle$ as the ancilla state for example and with the initial state well-approximated with fidelity $\geq 1-\delta$, then the final state $|\tilde{v}(t)\rangle$ is also well-approximated with fidelity $\geq 1-\delta$. When there are no errors in the implementation of the inverse Fourier transform, $|v(t)\rangle$ can similarly be well-approximated with fidelity $\geq 1-\delta$. \\

For the error in retrieving $|u(t)\rangle$, we take the heat equation as an example to demonstrate the error estimation method. Here $\partial u/\partial t=\nabla^2_x u$, which means the corresponding $w$ obeys 
\begin{align} \label{eq:wequation}
    \frac{\partial w}{\partial t}=-\nabla^2_x\frac{\partial w}{\partial \xi}
\end{align}
where $u(t,x)=\int_0^{\infty} w(t,x,\xi)d\xi$. In the Schr\"odingerisation framework, we choose $w(0,x, \xi)=\exp(-|\xi|)u(0,x)$, so we need the state $|\Xi\rangle$ for the quantum implementation. However, in the case where we start with a Gaussian state $|G\rangle$, we need a different initial condition 
\begin{align}
    w'(0,x,\xi)=\frac{1}{\sqrt{s}\pi^{1/4}}e^{-\xi^2/(2s^2)}u(0,x)
\end{align}
where $u(0,x)$ is still identical to the original scenario and only the ancilla mode changes. Here  $w'(t,x,\xi)$ still obeys the same PDE as $w(t,x,\xi)$ in Eq.~\eqref{eq:wequation}. Then we can define $u'(t,x)=\int_{0}^{\infty}w'(t,x,\xi)d\xi$. Now we want to estimate 
\begin{align}
    \|\vect{u}(t)-\vect{u}'(t)\|^2=\int_{-\infty}^{\infty}\left(\int_0^{\infty}(w(t,x,\xi)-w'(t,x,\xi))d\xi\right)^2 dx.
\end{align}
Both $w(0,x,\xi)$ and $w'(0,x,\xi)$ have exponential decay in $\xi>0$ and are waves moving to the left of the $\xi$ domain. This means there exists $\xi^*>0$, $\xi^*=O(1)$ such that $w \sim 0$ for $\xi>\xi^*$. This means we can replace $\int_0^{\infty} (w-w')^2d\xi \rightarrow \int_0^{\xi^*} (w-w')^2d\xi$. This implies 
\begin{align} \label{eq:u-u'}
    &  \|\vect{u}(t)-\vect{u}'(t)\|^2 \lesssim \int_{-\infty}^{\infty} \left(\int_0^{\xi^*}(w(t,x,\xi)-w'(t,x,\xi))d\xi\right)^2 dx \nonumber \\
    & \leq \int_{-\infty}^{\infty} \int_{0}^{\xi^*}d\xi' \int_0^{\xi^*}(w(t,x,\xi)-w'(t,x,\xi))^2 d\xi dx \leq \xi^*\|\vect{v}-\vect{v}'\|^2 
\end{align}
where the Cauchy-Schwartz inequality is used in the second line, and $\vect{v}'(t)=\iint_{-\infty}^{\infty}w'(t,x,\xi)|x\rangle |\xi\rangle dx d\xi$. We showed before $\|\vect{v}\|=\|\vect{u}(0)\|$. Using the normalised Gaussian ancilla state it is simple to see $\|\vect{v}'\|=\iint_{-\infty}^{\infty} w'(0,x,\xi)^2 d \xi dx=\|\vect{u}(0)\|=\|\vect{v}\|$. Therefore we can write 
\begin{align}
    1-\delta \leq |\langle v|v'\rangle|=1-\frac{1}{2\|\vect{u}(0)\|^2}\|\vect{v}-\vect{v}'\|^2. 
\end{align}
Inserting into Eq.~\eqref{eq:u-u'} we have the bound
\begin{align}
\|\vect{u}(t)-\vect{u}'(t)\|^2 \lesssim \xi^* \delta \|\vect{u}(0)\|^2.
\end{align}
The quantum fidelity between $|u(t)\rangle$ and $|u'(t)\rangle$ can be written 
\begin{align}
   & |\langle u(t)|u'(t)\rangle|=1-\frac{1}{2}\left\| \frac{\vect{u}(t)}{\|\vect{u}(t)\|}-\frac{\vect{u}'(t)}{\|\vect{u}'(t)\|}\right\|^2.
   \end{align}
   Now 
\begin{align}
    &\left\| \frac{\vect{u}(t)}{\|\vect{u}(t)\|}-\frac{\vect{u}'(t)}{\|\vect{u}'(t)\|}\right\|=\left\|\frac{\|\vect{u}'(t)\|\vect{u}(t)-\|\vect{u}(t)\| \vect{u}'(t)}{\|\vect{u}(t)\| \|\vect{u}'(t)\|}\right\|=\left\|\frac{\|\vect{u}'(t)\|(\vect{u}(t)-\vect{u}'(t))+(\|\vect{u}'(t)\|-\|\vect{u}(t)\|) \vect{u}'(t)}{\|\vect{u}(t)\| \|\vect{u}'(t)\|}\right\| \nonumber \\
    &\leq \frac{\|\vect{u}(t)-\vect{u}'(t)\|}{\|\vect{u}(t)\|}+\frac{\|\|\vect{u}'(t)\|-\|\vect{u}(t)\|\|}{\|\vect{u}(t)\|}\leq 2\frac{\|\vect{u}(t)-\vect{u}'(t)\|}{\|\vect{u}(t)\|}.
\end{align}
Therefore
\begin{align}
    |\langle u(t)|u'(t)\rangle| \geq 1-2\left(\frac{\|\vect{u}(t)-\vect{u}'(t)\|}{\|\vect{u}(t)\|}\right)^2\gtrsim 1-2\xi^* \delta \frac{\|\vect{u}(0)\|^2}{\|\vect{u}(t)\|^2}.
\end{align}
\section{Error estimates in Hamiltonian evolution} \label{app:robustness}
In the absence of error correction, analog quantum simulation is not conducted perfectly and there are errors in the parameters of the generating Hamiltonian during real implementation. It is then an important question to ask that, given a particular fixed final error in the state, how precise must one be in the tuning of the Hamiltonian parameters with respect to the dimension of the system? We study this question for some simple examples, first without Schr\"odingerisation then with Schr\"odingerisation.

\subsection{Simple convection equation example}
We have the following convection equation where $\alpha_j, \beta_j \in \mathbb{R}$ with the simple initial condition:
\begin{align}
    \frac{\partial u}{\partial t}-\sum_{j=1}^D \alpha_j \frac{\partial u}{\partial x_j}=0, \qquad u(0,x_1, ...x_D)=\prod_{j=1}^D \frac{\exp(-x_j^2/2)}{\pi^{1/4}}. 
\end{align}
Then defining
\begin{align}
    \vect{u}(t) \equiv \int u(t, x_1,...,x_D)|x_1,...,x_D\rangle dx_1...dx_D
\end{align}
where $\vect{u}(t)$ is already normalised so $\vect{u}(t)=|u(t)\rangle$, we have
\begin{align}
    \frac{d \vect{u}}{dt}=-i\vect{H} \vect{u}, \qquad \vect{H}=\vect{H}^{\dagger}
\end{align}
where the Hamiltonian 
\begin{align}
    \vect{H}=\sum_{j=1}^D \alpha_j \hat{p}_j
\end{align}
So to get the final state one applies the following unitary operations onto the initial state 
\begin{align}
    |u(t)\rangle=U_1...U_D |u(0)\rangle
\end{align}
where $U_j=\exp(-i \alpha_j t\hat{p}_j)$. In the absence of errors we want to final target state to be 
\begin{align}   |\text{target}\rangle=U_1|u_1(0)\rangle...U_D|u_D(0)\rangle
\end{align}
where for each $j=1,...,D$
\begin{align}
   |u_j(0)\rangle=\int \frac{e^{-x_j^2/2}}{\pi^{1/4}}|x_j\rangle dx_j. 
\end{align}
Suppose we have a constant error $\epsilon$ in each of the Hamiltonian parameters $\alpha_j$. Then the true state could instead be 
\begin{align}
    |\text{true}\rangle=V_1|u_1(0)\rangle...V_D|u_D(0)\rangle
\end{align}
where for each $j=1,...,D$
\begin{align}
    V_j=U_je^{-i \epsilon t \hat{p}_j}.
\end{align}
We can require the quantum fidelity between the target and the true state to be bound from below
\begin{align}
    |\langle \text{target}|\text{true}\rangle|=\prod_{j=1}^D|\langle u_j(0)|e^{-i\epsilon t\hat{p}_j}|u_j(0)\rangle| \geq 1-\Delta. 
\end{align}
Now
\begin{align}
    e^{-i\epsilon t\hat{p}_j}|u_j(0)\rangle=\int \frac{e^{-x_j^2}/2}{\pi^{1/4}}|x_j+\epsilon t\rangle dx_j
\end{align}
thus 
\begin{align}
    & \langle u_j(0)|e^{-i\epsilon t\hat{p}_j}|u_j(0)\rangle=\frac{1}{\sqrt{\pi}}\iint e^{-x_j^2/2} e^{(x'_j)^2/2}\delta(x'_j-x_j-\epsilon t) dx_j dx'_j \nonumber \\
    &=\frac{1}{\sqrt{\pi}}\int e^{-x_j^2/2}e^{-(x_j+\epsilon t)^2/2}dx_j=\frac{1}{\sqrt{\pi}}\int e^{-x_j^2-\epsilon t x_j -(\epsilon t)^2/2} dx_j=e^{-(\epsilon t)^2/4}.
\end{align}
To keep the fidelity bound for a constant $\Delta$
\begin{align}
    |\langle \text{target}|\text{true}\rangle|=e^{-(\epsilon t)^2 D/4} \geq 1-\Delta 
\end{align}
which implies it is sufficient to keep the error 
\begin{align}
    \epsilon \leq \frac{1}{\sqrt{D}} \frac{2}{t} \sqrt{\log \left(\frac{1}{1-\Delta}\right)}, 
\end{align}
so $\epsilon \geq O(1/\sqrt{D})$ and there is no exponential suppression of the error necessary with respect to the dimension in this case. \\

Something undesirable would be if $\epsilon \sim 1/e^D$, for example. It appears that so long as we have a product state and the single mode fidelities scale inverse exponentially with $\epsilon$ then we can always have $\epsilon \sim \text{poly}(1/\sqrt{D})$. 

\subsection{More general suggestive proof without Schr\"odingerisation}
Suppose we have the case of product states, where each qumode evolves separately with each Hamiltonian $\alpha_j\vect{H}_j$ where we take out the parameter $\alpha_j$ that is tuneable. We can pick a `worst-case' $\alpha_j\vect{H}_j$ and we only need to work out the fidelity with respect to this case only. We pick out the corresponding mode of the initial state $u_j(0)$ and expand it in the eigenbasis $\{|n\rangle\}$ of $\vect{H}_j$
\begin{align}
    |u_j(0)\rangle=\sum_{n=0}^{\infty}u_j^{(n)}(0)|n\rangle
\end{align}
and $\vect{H}_j|n\rangle=E_j^{(n)}|n\rangle$. Again, we assume there is a perturbation of order $\epsilon$ to each parameter $\alpha_j$. This means 
\begin{align}
    |\langle \text{target}|\text{true}\rangle|=\prod_{j=1}^D |\langle u_j(0)|e^{-i\epsilon t \vect{H}_j}|u_j(0)\rangle| \equiv \prod_{j=1}^D F_j.
\end{align}
Then 
\begin{align}
    \langle u_j(0)|e^{-i\epsilon t \vect{H}_j}|u_j(0)\rangle=\sum_{n=0}^{\infty} e^{-i\epsilon tE_j^{(n)}}\left|u_j^{(n)}\right|^2
\end{align}
and the absolute value squared gives
\begin{align} \label{eq:fj2}
     F_j^2 \equiv |\langle u_j(0)|e^{-i\epsilon t \vect{H}_j}|u_j(0)\rangle|^2=\left(\sum_{n=0}^{\infty} \cos(\epsilon t E_j^{(n)})|u_j^{(n)}|^2\right)^2+\left(\sum_{n=0}^{\infty} \sin(\epsilon t E_j^{(n)})|u_j^{(n)}|^2\right)^2.
\end{align}
Let's make the assumption that $\epsilon t E_j^{(n)} \ll 1$. Then we can Taylor expand the trigonometric terms. Using the following definitions of the normalisation of the initial state, the average energy of the initial state and standard deviation 
\begin{align}
    \sum_{n=0}^{\infty}|u_j^{(n)}|^2=1, \qquad \sum_{n=0}^{\infty}E_j^{(n)}|u_j^{(n)}|^2 \equiv \langle E_j\rangle, \qquad \sum_{n=0}^{\infty}(E_j^{(n)})^2|u_j^{(n)}|^2 \equiv \langle E^2_j\rangle, \qquad \delta E_j^2=\langle E^2_j\rangle-\langle E_j\rangle^2
\end{align}
we can write
\begin{align}
    F_j^2=1-(\epsilon t)^2\delta E_j^2+O((\epsilon t E_j)^3).
\end{align}
In the limit that we consider, this can be approximated to be 
\begin{align}
    F_j^2 \approx e^{-(\epsilon t)^2\delta E_j^2}
\end{align}
Therefore 
\begin{align}
     |\langle \text{target}|\text{true}\rangle| \geq e^{-\epsilon t D \delta E} \geq 1-\Delta
\end{align}
or 
\begin{align} \label{eq:errorHgeneral}
    \epsilon \leq \frac{1}{D}\frac{1}{t \delta E} \log \left(\frac{1}{1-\Delta}\right)
\end{align}
where $\delta E$ is the maximum standard deviation of the energy of any mode in the $D$-mode initial state. Here the energy of the state refers to the energy corresponding to the worst case Hamiltonian $\vect{H}_j$. Thus $\epsilon \sim O(1/D)$. \\


We note that in the case where the initial state is not a product state of single qumodes and the evolution involves entangling operations, the scaling of $\epsilon$ with respect $D$ is expected only to improve. This is because the total fidelity $|\langle \text{target}|\text{true}\rangle|$ would be a product of the fidelities of a number of quantum states that is \textit{smaller} than $D$. 

\subsection{Error estimate using Schr\"odingerisation} 
The robustness analysis of the above holds in the same way for the step in Schr\"odingerisation involving $|\tilde{\vect{v}}(t)\rangle$. The only extra step in Schr\"odingerisation involves making a projective measurement $\hat{P}$ onto the ancillary mode, and here the error will then have a component dependent on the ratio of the norms of the states $\|\vect{u}(0)\|/\|\vect{u}(t)\|$. \\

In this case, using the same proof method as Appendix~\ref{app:initialstatepreparation}, if the target final state is $\vect{u}(t)_{target}$ and the true final state (with errors) is $\vect{u}(t)_{true}$, then 
\begin{align} \label{eq:ufidelity}
    |\langle u(t)_{target}|u(t)_{true}\rangle| \gtrsim 1-2\xi^* \Delta \frac{\|\vect{u}(0)\|^2}{\|\vect{u}(t)_{target}\|^2}
\end{align}
where 
\begin{align}
   |\langle \text{target}|\text{true}\rangle|= |\langle v_{target}|v_{true}\rangle| \geq 1- \Delta
\end{align}
and $\xi^*$ is the $\xi^*=O(1)>0$ value where $w \sim 0$ for $\xi>\xi^*$. Now suppose we have a condition
\begin{align}
    |\langle u(t)_{target}|u(t)_{true}\rangle|  \gtrsim 1-2\xi^* \Delta \frac{\|\vect{u}(0)\|^2}{\|\vect{u}(t)_{target}\|^2} \geq 1-\eta
\end{align}
where $1-\eta$ is a sufficient lower bound to the quantum fidelity between the final target state and true state for the original problem (where Schr\"odingerisation is necessary). This means that we can write
\begin{align}
    \Delta \lesssim \frac{1}{2\xi^*}\eta \frac{\|\vect{u}(t)_{target}\|^2}{\|\vect{u}(0)\|^2} \equiv \Delta^*
\end{align}
Here $\xi^*$ is a constant and $1-\eta$ is lower bound to quantum fidelity of the target and the true state. Inserting into Eq.~\eqref{eq:errorHgeneral} we find that the required error $\epsilon$ in controlling the parameters of $\vect{H}$ scale like
\begin{align}
    \epsilon \lesssim \frac{1}{D}\frac{1}{t \delta E}\log\left(\frac{1}{1-\Delta^*}\right).
\end{align} 

\end{document}